\documentclass[10pt,journal,compsoc]{ieeetran}

\usepackage{listings}
\usepackage{hyperref}
\usepackage{algorithm}
\usepackage{algorithmicx}
\usepackage[noend]{algpseudocode}
\usepackage{varwidth}
\usepackage{amsmath}
\usepackage{graphicx}
\usepackage{multirow}
\usepackage{colortbl}
\usepackage[x11names]{xcolor}

\hypersetup{
    colorlinks=true,
    linkcolor=blue,
    filecolor=magenta,      
    urlcolor=blue,
}

\algnewcommand\algorithmicInput{\textbf{Input:}}
\algnewcommand\Input{\item[\algorithmicInput]}

\algnewcommand\algorithmicOutput{\textbf{Output:}}
\algnewcommand\Output{\item[\algorithmicOutput]}

\newcommand{\code}[1]{\text{\lstinline[basicstyle=\ttfamily\small]~#1~}}
\newcommand{\Pair}[2]{\mbox{$\langle #1,#2 \rangle$}}

\definecolor{javared}{rgb}{0.6,0,0} 
\definecolor{javagreen}{rgb}{0.25,0.5,0.35} 
\definecolor{javapurple}{rgb}{0.5,0,0.35} 
\definecolor{javadocblue}{rgb}{0.25,0.35,0.75} 

\lstdefinelanguage{JavaScriptColor}{
  keywords={break, case, catch, const, continue, debugger, default, delete, do, else,
    finally, for, function, if, in, instanceof, new, return, switch, this,
    throw, try, typeof, var, void, while, with,
    class, interface, implements, public, private, constructor
    },
  morecomment=[l]{//},
  morecomment=[s]{/*}{*/},
  morestring=[b]',
  morestring=[b]",
  keywordstyle=\color{javapurple}\bfseries,
  identifierstyle=\color{black},
  commentstyle=\color{javagreen}\ttfamily,
  numberstyle=\color{javared}\ttfamily,
  stringstyle=\color{javared}\ttfamily,
  sensitive=false,
  numbers=left,
  stepnumber=1,
  escapeinside={/*\#}{\#*/},
}
\newcommand{\WRAP}[2]{\begin{minipage}[t]{#1}{#2}\end{minipage}}

\newcommand{\TotalProjects}{127,531}
\newcommand{\TotalProjectsApprox}{127,500}
\newcommand{\ProjectsUsingEvents}{35,757}
\newcommand{\TotalPairs}{532,004}
\newcommand{\TotalPairsApprox}{532,000}
\newcommand{\TotalUniquePairs}{160,195}
\newcommand{\TotalIncorrectPairs}{75}
\newcommand{\BestTPRate}{90.9}
\newcommand{\BestRecallRate}{7.5}
\newcommand{\TotalLabeledCorrect}{959}
\newcommand{\TotalLabeledIncorrect}{399}
\newcommand{\TotalLabeledImprecise}{4,323}

\newcommand{\TotalLabeledPairs}{5,681}
\newcommand{\NumAnalyzedProjects}{25}
\newcommand{\NumReportedIssues}{30}
\newcommand{\NumConfirmedBugs}{7}

\newcommand{\LearnedCorrectTerm}{expected}
\newcommand{\LearnedIncorrectTerm}{anomalous}
\newcommand{\ValidationSetCorrectTerm}{correct}
\newcommand{\ValidationSetIncorrectTerm}{incorrect}
\newcommand{\ValidationSetImpreciseTerm}{imprecise}

\begin{document}

\title{Learning How to Listen: Automatically Finding Bug Patterns in Event-Driven JavaScript APIs}

\author{Ellen Arteca, Max Sch{\"{a}}fer, Frank Tip
  \IEEEcompsocitemizethanks{
    \IEEEcompsocthanksitem E. Arteca and F. Tip are with the Khoury College of Computer Sciences at Northeastern University. E-mail: \{arteca.e,f.tip\}@northeastern.edu
    \IEEEcompsocthanksitem M. Sch{\"{a}}fer is with GitHub. E-mail: max-schaefer@github.com
  }
}

\IEEEpubid{}

\maketitle

\begin{abstract}
  Event-driven programming is widely practiced in the JavaScript community, both on
  the client side to handle UI events and AJAX requests, and on the server
  side to accommodate long-running operations such as file or network I/O.
  Many popular event-based APIs allow event names to be specified as free-form strings without any validation,
  potentially leading to \textit{lost events} for which no listener has
  been registered and \textit{dead listeners} for events that are never emitted.
  In previous work, Madsen et al. presented a precise static analysis for detecting such problems,
  but their analysis does not scale because it may require a number of contexts that is exponential in the size of the program.
  Concentrating on the problem of detecting dead listeners, we present an approach to \textit{learn}
  how to use event-based APIs by first mining a large corpus of JavaScript code using a simple
  static analysis to identify code snippets that register an event listener, and then applying statistical
  modeling to identify \LearnedIncorrectTerm{} patterns, which often indicate \ValidationSetIncorrectTerm{} API usage.
  In a large-scale evaluation on \TotalProjects{} open-source JavaScript code bases, our technique was able to detect \TotalIncorrectPairs{} \LearnedIncorrectTerm{} listener-registration patterns, while maintaining a precision of \BestTPRate{}\% and recall of \BestRecallRate{}\% over a validation set, demonstrating that a learning-based approach to
  detecting event-handling bug patterns is feasible.  
  In an additional experiment, we investigated instances of these patterns in \NumAnalyzedProjects{} open-source projects,
  and reported \NumReportedIssues{} issues to the project maintainers, of which
  \NumConfirmedBugs{} have been confirmed as bugs.  
\end{abstract}

\begin{IEEEkeywords}
   static analysis, JavaScript, event-driven programming, bug finding, API modeling
\end{IEEEkeywords}

\section{Introduction}\label{IntroductionSection}

Event-driven programming has been the dominant paradigm in JavaScript since its early days.
This is quite natural on the client side, since most web applications are GUI-based and hence are centered around reacting to user actions such as clicking a button or pressing a key.
The W3C UI Events standard \cite{W3CEvents} defines the low-level event API supported by all modern browsers, while popular libraries such as jQuery \cite{jQuery}, Angular \cite{Angular} and React \cite{React} provide higher-level abstractions on top of it.
Many other client-side APIs such as Web Workers and Web Sockets are likewise programmed in an event-driven style.
On the desktop, the popular Electron \cite{Electron} framework enforces an architecture where applications are split into a main process and a renderer process, which communicate via an event-based API.
Finally, the Node.js platform \cite{NodeJS}, which is dominant in server-side JavaScript, advocates an asynchronous programming style centered around a collection of event-based APIs for accessing resources like the file system, the network, or databases.

The precise APIs implemented by individual platforms and frameworks differ, but a common feature across all of JavaScript is the notion of a central \emph{event loop} that handles event dispatching.
Events are identified by an \emph{event name} and may optionally have a payload.
When an event happens, it is associated with a particular object, which is known as the \emph{event target} in many client-side frameworks and the \emph{event emitter} in Node.js.
We will follow the latter terminology in this paper.
Client code can register \emph{listener functions} (or \emph{listeners} for short) for a particular event on an event emitter.
When an event is emitted, all the listener callbacks registered for it on the emitter object are run in sequence.
While many events are emitted by framework code, application code can also emit events explicitly.

Most of the event-based APIs mentioned above are intrinsically \emph{dynamic} and \emph{untyped}.
By ``dynamic'' we mean that the association between events and listeners can change over time, with new listeners being registered and existing listeners being removed throughout an event emitter's lifecycle.
Indeed, it is common for listeners themselves to register or remove listeners on their own or on other emitter(s).
By ``untyped'' we mean that event names are free-form strings that are not validated in any way, and can be associated with any emitter and any payload.
In particular, client applications can emit and listen for custom events on emitters defined by a library.

\begin{figure*}
   \centering
       \includegraphics[width=.65\textwidth]{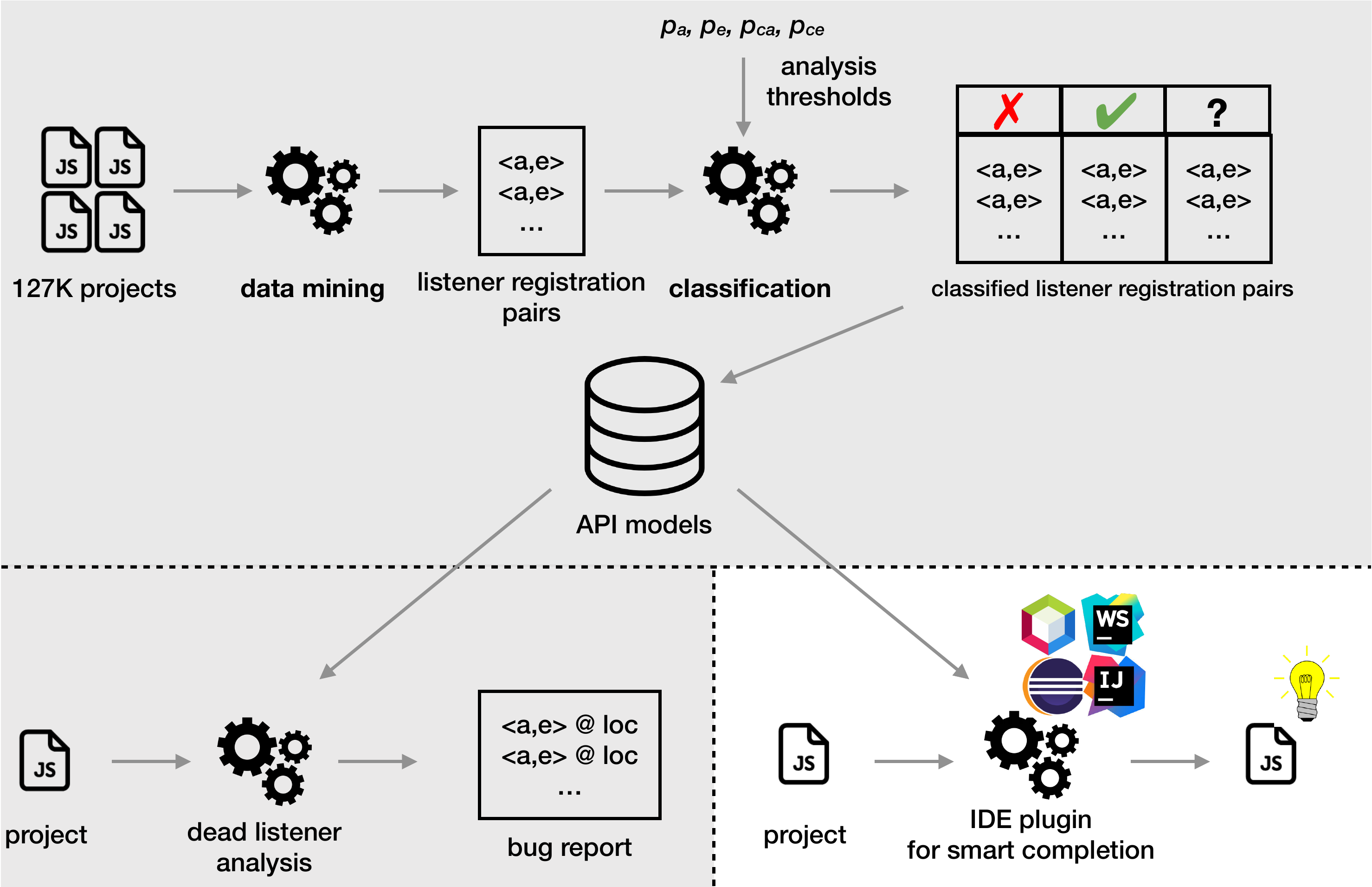}
 \caption{Overview of approach: the top half depicts the model-construction pipeline, while the bottom half shows their potential applications. This paper focuses on the shaded areas.}\label{fig:pipelineDiag}
\end{figure*}

While these two properties are prized by some for their flexibility, they also give rise to several classes of subtle bugs~\cite{DBLP:conf/oopsla/MadsenTL15}.
For example, if a listener registration misspells the name of the event or registers the listener on the wrong object, the listener will never be invoked.
This is known as a \emph{dead listener}.
Dead listeners can also arise if a listener is registered at the wrong time, for instance after the event has already been emitted.
The dual of a dead listener is a \emph{lost event}, which can happen if an event emission misspells the event name or emits it on the wrong object.
Both dead listeners and lost events are particularly hard to debug, as they manifest in the lack of execution of the listener function rather than an explicit error message.

In this paper, we concentrate mostly on dead-listener bugs.
Our goal is to detect such bugs \emph{automatically} and \emph{statically}, i.e., without having to run the code under analysis.

Prior work by Madsen et al.~\cite{DBLP:conf/oopsla/MadsenTL15} employs context-sensitive static analysis techniques to infer a semantic model of event emission and listener registration to identify dead listeners.
Unfortunately, their analysis does not scale well because it may require a number of contexts that is exponential in the size of the program.

We propose instead to \textit{learn} how to use event-based APIs by first mining a large corpus of JavaScript code with a simple static analysis to identify code snippets that register an event listener, and then applying statistical modeling to identify {\it \LearnedIncorrectTerm{} patterns}.
Intuitively, if we look at enough code we would expect most API usages to correspond to their designed use, so particularly rare patterns are likely bugs.
We formalize this concept of ``particularly rare'' as thresholds in our statistical analysis, and identify patterns that meet these thresholds as potential bugs.
Using the same thresholds, our approach also addresses the dual problem of learning \emph{\LearnedCorrectTerm{}} uses, with ``particularly common'' uses of the APIs corresponding to the intended use.

Figure~\ref{fig:pipelineDiag} visualizes our approach.
The top of the figure shows how models of event-driven APIs are constructed in two steps:
First, a \emph{data mining} analysis is applied to a large number of JavaScript projects to obtain a list of 
 event listener registrations. These are represented as 
\emph{listener-registration pairs} $\Pair{a}{e}$, where $a$
is an event-emitting API endpoint symbolically represented by an \emph{access path}~\cite{mezzetti18} as explained in Section \ref{DataMiningSection}, and $e$ is the name of the event the listener is registered for.
The second step is \emph{classification}, i.e., performing a statistical analysis 
of the occurrence distributions of $e$'s and $a$'s, and using this to identify pairs $\Pair{a}{e}$ where the access path $a$ and event $e$ are rare relative to each other.
In other words, we look for cases where $e$ is rarely listened for on $a$, and $a$ rarely registers a listener for $e$. 

Considering one of these conditions in isolation, or only the absolute number of occurrences of a pair, is not usually sufficient, since the data may be too sparse to conclude that it is \LearnedIncorrectTerm{}.
For example, $a$ may be a rarely-used API, or $e$ may be a custom event that is used only by one particular code base.
If, however, both the event emitter and the event name are rare for each other yet otherwise common, then that is a strong indication that this pair represents a mistake.

Our statistical analysis has four parameters shown as inputs to the classification stage in Figure \ref{fig:pipelineDiag}: 
rarity thresholds $p_a$ and $p_e$ defining when paths and events are considered rare, respectively, 
and confidence thresholds $p_{\mbox{\it ca}}$ and $p_{\mbox{\it ce}}$ defining the statistical confidence we demand for paths and events to be considered rare, respectively.
The output of classification is a set of pairs learned to be {\it \LearnedCorrectTerm{}}, 
and a set learned to be {\it \LearnedIncorrectTerm{}}.
Pairs are left unclassified if they do not meet the thresholds for being common or rare.

These sets constitute API models, for those APIs analyzed.
Once constructed, these API models can be used, e.g., in bug finding tools 
(see bottom left part of Figure~\ref{fig:pipelineDiag}), or for smart completion in an IDE
(see bottom right part).
In this work, we focus on the set of pairs that are learned to be \LearnedIncorrectTerm{}, as they are likely to indicate dead listener bugs.

The effectiveness of our approach crucially depends on how we configure the threshold parameters for classification.
In our evaluation, we systematically explore the space of possible configurations, computing for each of them the set of \LearnedIncorrectTerm{} listener-registration pairs from more than \TotalPairsApprox{} pairs mined from over \TotalProjectsApprox{} open-source code bases.
To quantitatively assess the quality of the models generated with a particular configuration, we then compute the true-positive rate (the \emph{precision}) and the percentage of true positives detected (the \emph{recall}) with respect to a \textit{validation set} of pairs that we semi-automatically 
labeled as {\it \ValidationSetCorrectTerm{}} or {\it \ValidationSetIncorrectTerm{} according to the API documentation}%
\footnote{
  Event-listener registration pairs in the validation set may also be designated as being \ValidationSetImpreciseTerm{}, to reflect situations
  where the access path is insufficiently precise to make a determination (see Section~\ref{MethodologySection}).
}.

In general, configurations with lower precision yield higher recall.
For practically useful tools, however, a precision of at least 90\% is generally considered essential~\cite{Sadowski18,Sadowski15}.
Several configurations achieve this rate over the labeled set.

To gain confidence that this is not simply an artifact of the data, we performed a 10-fold cross-validation experiment.
We partitioned the labeled set into 10 sets; for each set, we found the optimal configuration for the other 9 sets 
(which together form the training data), and computed the precision and recall of that configuration over the remaining set (which comprises
the validation data).
Our results show that the optimal configuration for the training data consistently achieves good results over the validation data.

To qualitatively assess the usefulness of our approach, we investigated uses of anomalous pairs
in \NumAnalyzedProjects{} open-source projects, reporting \NumReportedIssues{} issues to the project maintainers.
At the time of writing, \NumConfirmedBugs{} of these have been confirmed as dead-listener bugs, and two have been patched.

The rest of the paper is structured as follows.
Section~\ref{BackgroundSection} provides background on event-driven JavaScript programming  and reviews a dead-listener bug in an open-source project.
Sections~\ref{DataMiningSection} and~\ref{ClassificationSection} explain our approach in detail, while Section~\ref{ImplementationSection} covers the implementation.
Sections~\ref{MethodologySection} covers experimental methodology used in an experimental evaluation that is presented
in Section~\ref{EvalSection}. Next, Section~\ref{CaseStudyAndThreatsSection} presents a case study of false positives and false negatives
observed in our results, and discusses threats to validity.
Section~\ref{LostEventsSection} discusses to what extent our techniques are applicable to detecting lost events. 
Section~\ref{RelatedWorkSection} reviews related work, and Section~\ref{ConclusionSection} concludes and outlines directions for future work.

\begin{sloppypar}
The source code of our implementation, experimental data, and reproduction instructions are available online at \href{https://github.com/emarteca/JSEventAPIModelling}{\textcolor{blue}{https://github.com/emarteca/JSEventAPIModelling}}
\end{sloppypar}

\section{Background}\label{BackgroundSection}

We begin by recapitulating the basics of event-driven programming in Node.js and some of the most common kinds of mistakes programmers make when writing event-driven code. We then show a concrete example of such a bug, based on code we found using our approach in an open-source project on GitHub, and finally explain how we go about identifying this sort of bug automatically.

\subsection{Event-driven programming in Node.js}
All event emitters in Node.js are instances of the \code{EventEmitter} class~\cite{EventEmitterDocs} or one of its subclasses. 
Listeners are associated with an event by invoking one of several listener registration methods (such as \code{on} or \code{addListener}); these all take two arguments: an event name, which is a free-form string, and the listener function itself. 
Events can be emitted by invoking the \code{emit} method, which takes as its first argument an event name; any further arguments are passed as arguments to the listener functions associated with the event.

\begin{sloppypar}
  A typical example of this event-driven style is the \code{request} function from the \code{http} package in the Node.js standard library. Normally invoked as \code{http.request(url, fn)} where \code{url} is the URL to make a request to, and \code{fn} is a listener function, it creates an event emitter object of class \code{http.ClientRequest} representing the pending request to \code{url} and associates \code{fn} with the \code{response} event of the request.
\end{sloppypar}

When a response to the request is received, the \code{response} event is emitted, causing \code{fn} to be invoked with an argument that is an instance of \code{http.IncomingMessage} representing the HTTP response. This object is itself an event emitter, emitting \code{data} events when response data becomes available and an \code{end} event once all data has been received.

If, on the other hand, the request times out before receiving a response, the request object emits a \code{timeout} event.

\subsection{Motivating example}
Consider the code shown in Figure~\ref{fig:running-example}, which is a condensed version of a bug our approach identified in the \href{https://www.npmjs.com/package/min-req-promise}{\texttt{min-req-promise}} npm package.

\begin{figure}
{\small
\begin{lstlisting}
const http = require('http');
module.exports.request = (url) =>
  new Promise((resolve, reject) => {
    const req = http.request(url, res => { /*#\label{line:res}#*/
      res.on('data', /* omitted */); /*#\label{line:res-data}#*/
      res.on('end', () => { /*#\label{line:res-end}#*/
          /* omitted */ 
          resolve( res);
      }); 
      res.on('timeout', () => reject(req)); /*#\label{line:res-timeout}#*/// bug here
    });                                     
    req.end(); /*#\label{line:req-end}#*/  
  });
\end{lstlisting} 
}
\vspace*{-5mm}
\caption{An example of a dead-listener bug}\label{fig:running-example}
\end{figure}

\begin{sloppypar}
\texttt{min-req-promise}  turns the somewhat intricate event-based \code{http.request} API discussed 
above into a simpler promise-based API. It exports a function \code{request}, which returns a promise wrapped around a call to \code{http.request}. The pending request (an instance of \code{http.ClientRequest}) is stored in variable \code{req} (line~\ref{line:res}), and a listener function is passed to \code{http.request} on the same line, which associates it with the \code{response} event on \code{req}. Finally, \code{req.end()} is called on line~\ref{line:req-end} to dispatch the request. Once a response arrives, the \code{http} library invokes the listener provided on line~\ref{line:res}, passing it a \code{res} object representing the response, which is an instance of \code{http.IncomingMessage}. On this object, handlers for three events are installed: \code{data}, \code{end} and \code{timeout}. The first event is emitted whenever a chunk of response data arrives, the second when the response has been received in its entirety. For simplicity, we have omitted the handler functions for these two events; the interested reader is referred to the project's GitHub page \cite{MinReqPromise}.
\end{sloppypar}

\begin{sloppypar}
The third event, \code{timeout}, is the problematic one: this event is actually never emitted by \code{http.IncomingMessage} objects, so the listener on line~\ref{line:res-timeout} is dead code. There is a \code{timeout} event on \code{http.ClientRequest}, however, so presumably the event should have been registered on \code{req}, not \code{res}. We contacted the author of \texttt{min-req-promise}, who confirmed our analysis of the issue.
\end{sloppypar}

Note that there are no compile-time or runtime diagnostics to alert the developer to this problem: not only is it very difficult to infer precise types for variables in JavaScript in general, but there is not even anything semantically wrong with registering a handler for a \code{timeout} event on \code{http.ClientRequest}. While the \code{http} library will never emit this event, client code could do so itself by calling the \code{emit} method (although in this case it does not). 
Moreover, since dead-listener bugs do not cause a crash at runtime, they may go undetected for a long time: in the case of \code{min-req-promise}, the bug had been present since its initial version (released in March 2018).

At present, the only way for a developer to detect this sort of problem is to carefully reason about types and the events they support (as we have done above), or to write extensive unit tests to ensure all events are handled as expected. In the above example, this would require adding a test involving a request that times out, which is an edge case that is easy to overlook.

Clearly, a more automated approach is desirable.

\subsection{Automatically detecting dead listeners}

We have argued that the dynamic nature of the JavaScript event-driven APIs makes it unrealistic to detect dead listeners at runtime. However, an approach based on static analysis faces the usual dilemma of having to trade off precision against performance: an imprecise analysis is likely to report many false positives, while a very precise analysis will not usually scale to realistic code bases.

Ideally, a static analysis would analyze client code as in Figure~\ref{fig:running-example} along with the implementation of the Node.js standard libraries and any other third-party libraries it depends on, derive a precise model of which types support which events, and then flag dead listeners based on this information.
In practice, we know of no static analyzer for JavaScript precise enough to derive such a model that scales to the size and complexity of the libraries involved. As a comparatively benign example, the Node.js \code{http} package transitively depends on more than 60 modules, for a total of around 20,000 lines of code. While this is quite manageable for, say, type inference or taint tracking, it is out of reach for techniques that precisely model event dispatch, such as that of Madsen et al.~\cite{DBLP:conf/oopsla/MadsenTL15}.

The usual answer is to instead provide the analysis with simplified models of the libraries involved. This is indeed a good approach for frequently used and well-documented packages like \code{http}, but the modern JavaScript library landscape is vast, with npm alone hosting well over one million packages. While many of these are very rarely used, the number of popular packages is still too large to allow manual modeling, especially since packages tend to go in and out of style quite frequently.

\subsection{Approach}\label{ApproachSection}

Our proposed solution to this dilemma is to turn the size of the JavaScript ecosystem to our advantage in a two-step approach illustrated in Figure~\ref{fig:pipelineDiag}: first, we mine large amounts of open-source code from GitHub and other hosting platforms for real-world examples of event-listener registrations; then we perform a statistical analysis to determine whether a certain pattern is rare and hence suggestive of \ValidationSetIncorrectTerm{} API usage, or whether is common and therefore likely to be a \ValidationSetCorrectTerm{} use. This allows us to automatically derive models instead of writing them by hand.

In the next two sections we explain the data mining and classification steps in more detail.

\section{Data Mining}\label{DataMiningSection}

The mining step is implemented as a context- and flow-insensitive static analysis that finds event-listener registrations and records them as listener-registration pairs of the form $\Pair{a}{e}$ where $a$ represents the object on which the listener is registered, and $e$ the event for which it is registered.

Both $a$ and $e$ need to be represented in a code base-independent way to enable us to meaningfully collate results obtained on many different code bases.

For events, this is easy: $e$ is the event name annotated with the emitter package. For instance, {\tt timeout} events on $a$'s rooted in the {\tt http} package are considered to be different from {\tt timeout} events rooted in the {\tt process} package.
This is important, as events with the same name in different packages may behave differently.

To represent event emitters, we use access paths similar to those proposed by Mezzetti et al.~\cite{mezzetti18}: starting from a package import, the access path records a sequence of property reads, method calls and function parameters that need to be traversed to reach a particular point in the program. 
More precisely, $a$ conforms to the following grammar:
\[\begin{array}{lcll}
  a & ::= & \mathbf{require}(m) & \WRAP{50mm}{an import of package $m$} \\
  & \mid & a.f & \WRAP{50mm}{property $f$ of an object represented by $a$} \\
  & \mid & a() & \WRAP{50mm}{return value of a function represented by $a$} \\
  & \mid & a(i) & \WRAP{50mm}{$i$th argument of a function represented by $a$} \\
  & \mid & a_{\mathbf{new}}() & \WRAP{50mm}{instance of a class represented by $a$}
\end{array}\]

Note that access paths are always rooted at a package import, so we can always tell which package any program element derives from.

For instance, in Figure~\ref{fig:running-example}, the access path associated with the variable \code{req} is $\mathbf{require}(\mathtt{http}).\mathtt{request}()$, meaning that \code{req} is initialized to the result of calling the method \code{request} on the result of importing the \code{http} module.\footnote{Note that the argument to  \code{request} is not recorded in the access path; see also Section~\ref{CaseStudyAndThreatsSection}.}

\begin{sloppypar}
The access path of \code{res}, on the other hand, is $\mathbf{require}(\mathtt{http}).\mathtt{request}(1)(0)$: starting from the import of \code{http}, we look at a call to \code{request} as above, but instead of considering the result we look instead at its second argument,~\footnote{
We index arguments starting from zero, so the argument at index one is the second argument.
}
which is the listener function on line~\ref{line:res}, and then the first argument to that function, which is the variable \code{res}. 
As above, the value of the first argument to \code{request} is not recorded in the access path.
\end{sloppypar}

Upon analyzing this snippet of code, we would record three pairs of access paths and events, corresponding to the three explicit event listener registrations:

\begin{enumerate}
\item $\Pair{\mathbf{require}(\mathtt{http}).\mathtt{request}(1)(0)}{\mathtt{data}}$, corresponds to line~\ref{line:res-data}
\item $\Pair{\mathbf{require}(\mathtt{http}).\mathtt{request}(1)(0)}{\mathtt{end}}$, corresponds to line~\ref{line:res-end}
\item $\Pair{\mathbf{require}(\mathtt{http}).\mathtt{request}(1)(0)}{\mathtt{timeout}}$, corresponds to line~\ref{line:res-timeout}
\end{enumerate}

Our approach is based on the assumption that if such pairs are collected over a lot of code, we are likely to see many instances of the first two (\ValidationSetCorrectTerm{}) pairs, but few instances of the last (\ValidationSetIncorrectTerm{}) pair. 
This is indeed the case: in our experiments (further detailed below) we found 996 instances of the first pair and 898 of the second, but only one of the third.

To detect event-listener registrations, our analysis looks for calls to methods named {\tt on}, {\tt once}, {\tt addListener}, {\tt prependOnceListener} or {\tt prependListener} (the standard Node.js listener registration methods), where the receiver can be represented by an access path, the first argument is a constant string (the event name), and the second argument is a function (the callback).

\subsection{Access Path Imprecision}\label{ImprecisionSection}
Due to its simplicity, our mining analysis is fairly imprecise. As we will show in Section~\ref{EvalSection}, this does not matter: the statistical analysis in the classification step compensates for much of the imprecision and yields high-quality results.
There are two main sources of imprecision: our choice of access paths to represent runtime objects, and the lack of context and flow sensitivity of the analysis.

\subsubsection{Imprecision due to access path representation}\label{ImpreciseAccessPathsSection}
The formulation of access paths we use is attractive in its simplicity, but it is imprecise because access paths are both \emph{overapproximate} (the same access path may represent many different runtime objects) and \emph{non-canonical} (two different access paths may represent the same runtime object).

As an example of the former, consider again line~\ref{line:res} in Figure~\ref{fig:running-example}. This line can equivalently be written like this:

\begin{lstlisting}
const req = http.request(url);
req.on('response', res => { ... });
\end{lstlisting}

\begin{sloppypar}

Here, the access path for {\tt res} becomes $\mathbf{require}(\mathtt{http}).\mathtt{request}().\mathtt{on}(1)(0)$: it is the first parameter of the second argument to \code{on} invoked on the result of \code{http.request}. 
This does \emph{not} record the other arguments to \code{on}; so, the access path does not include the fact that the first argument to \code{on} is \code{response}. 
While in actual fact \code{res} is an instance of \code{http.IncomingMessage} since the event listener is associated with event \code{response}, the parameter of an event listener associated with, for example, event \code{socket} has the same access path, but it is an instance of \code{net.Socket}. 
This means that in some cases we cannot determine event registration correctness based purely on the object's access path: for example, while both \code{http.IncomingMessage} and \code{net.Socket} have a \code{data} event, the former has an \code{aborted} event that the latter lacks. 
\end{sloppypar}

As an example of the lack of canonicity of access paths, note that the event registration method \code{on} returns the emitter event on which it is invoked, so lines~\ref{line:res-data}--\ref{line:res-timeout} of Figure~\ref{fig:running-example} could be rewritten as a single statement with three chained listener registrations:

\begin{lstlisting}
res.on('data', /* omitted */)
   .on('end', /* omitted */)
   .on('timeout', () => reject(req));
\end{lstlisting}

\begin{sloppypar}
  While this does not affect the pair recorded for the first registration, the second becomes $\Pair{\mathbf{require}(\mathtt{http}).\mathtt{request}(1)(0).\mathtt{on}()}{\mathtt{end}}$. 
Semantically, $\mathbf{require}(\mathtt{http}).\mathtt{request}(1)(0).\mathtt{on}()$ and $\mathbf{require}(\mathtt{http}).\mathtt{request}(1)(0)$ denote the same set of concrete runtime objects, i.e., they are \emph{aliases}.
\end{sloppypar}

\subsubsection{Access path alias removal}\label{sec:APAliasRemoval}

Such \emph{chained listener registrations} are a very common pattern in event-driven JavaScript, since all listener-registration methods return their receiver
object, that is, the event emitter object itself. To mitigate the resulting aliasing, we replace access paths representing the result of a listener-registration method
with the access path of the receiver object of the call.

\begin{sloppypar}
For example, recall that the access path for \code{res} in Figure~\ref{fig:running-example} is $\mathbf{require}(\mathtt{http}).\mathtt{request}(1)(0)$.
This means that the access path for \code{res.on('data', ...)} is $\mathbf{require}(\mathtt{http}).\mathtt{request}(1)(0).\mathtt{on}()$.
Similarly, the access path for \code{res.on('data', ...).on('end', ...)} is $\mathbf{require}(\mathtt{http}).\mathtt{request}(1)(0).\mathtt{on}().\mathtt{on}()$.
However, the latter two access paths are aliases of the first one, so we replace both of them with $\mathbf{require}(\mathtt{http}).\mathtt{request}(1)(0)$,
which enables our analysis to recognize that all three event listeners are registered on the same API element.
\end{sloppypar}

Note that in general, cycles in the data-flow graph can give rise to infinitely many access paths that all alias each other. Such cycles are already 
detected and collapsed by the access-path library used in our implementation.

\subsubsection{Imprecision due to lack of context and flow sensitivity}\label{AnalysisImprecisionSection}
The second source of imprecision is the lack of context and flow sensitivity of the analysis, which may cause listener-registration pairs to be reported that can never actually happen at runtime.

\begin{figure}
\begin{lstlisting}
var eos = function(stream, opts, callback) {
  // ...
  if ( isRequest( stream)) {
    stream.on('complete', /* ... */ );
    stream.on('abort', /* ... */);
  }
  // ...
\end{lstlisting}
\vspace*{-5mm}
\caption{Listener registration with explicit type check}\label{fig:typeReflectionExample}
\end{figure}

A typical example of this is shown in Figure~\ref{fig:typeReflectionExample}.\footnote{Adapted from the \href{https://github.com/mafintosh/end-of-stream}{\color{blue}{{\tt mafintosh/end-of-stream} project on GitHub}}}
The function \code{eos} accepts a variety of streams. Since the {\tt complete} and {\tt abort} events are not emitted by all types of streams,
it first checks whether the stream is a request before registering listeners for these two events.
Our analysis lacks flow sensitivity, and hence reports \code{complete} and \code{abort} event listeners being registered on \emph{all} streams passed as arguments to \code{eos}
that do \emph{not} support these events (in this particular example, stream objects of class {\tt http.IncomingMessage}). 

Finally, note that our mining analysis does not account for code that explicitly emits an event. This means that
it may report a pair $\Pair{a}{e}$ that is, in general, \ValidationSetIncorrectTerm{} because $a$ is a library API that does not emit event
$e$, but happens to be a \ValidationSetCorrectTerm{} use for a particular code base, because that code base explicitly emits $e$ on $a$.

\begin{figure}
\begin{lstlisting}
var http = require('http');
var server = http.createServer();
var client = http.request();
client.on('response', function(rsp) {
    server.emit('response', /* ... */) /*#\label{line:manual-emit}#*/
});
server.once('response', 
            function(data) { /* ... */ }); /*#\label{line:listen-response}#*/
\end{lstlisting}
\vspace*{-5mm}
\caption{Explicit emission of non-standard event}\label{fig:manualEmitExample}
\end{figure}

For example, consider Figure \ref{fig:manualEmitExample}.\footnote{Adapted from the \href{https://github.com/strongloop/strong-pm}{\color{blue}{{\tt strong\-loop/strong-pm} project on GitHub}}}
On line~\ref{line:listen-response}, we see a listener to {\tt response} registered on the result of a call to {\tt http.createServer()}, which is an object that is an instance of {\tt http.Server}.
According to the API documentation of the \code{http} library, \code{http.Server} does not emit the \code{response} event.
However, on line~\ref{line:manual-emit}, the client application itself emits a {\tt response} on the server object, and hence the \code{response} listener is not dead.
The analysis could be improved to suppress listener-registration pairs for which it sees an explicit emit, but we decided against doing this in the interest of simplicity.

These various sources of imprecision can make the data produced by the mining step somewhat noisy, but the classification step mitigates this problem: its input is collected from a large set of code bases, the majority of which do not use intricate idioms like these.

\section{Classification}\label{ClassificationSection}

Once we have collected a large corpus of listener-registration pairs we want to classify them to identify pairs that are likely to correspond to API misuses. We first describe the general intuition behind the approach, which we make precise in a statistical analysis. We then explain how this analysis is applied to classify pairs as \LearnedIncorrectTerm{}, and finally introduce a refinement to avoid misclassifications.

\subsection{General intuition}\label{ModelGeneralIntuitionSec}
\begin{sloppypar}
As argued above, if the analyzed corpus is big enough, buggy listener-registration pairs are likely to be relatively rare.
However, the converse is not true: there are at least two cases where a rare listener-registration pair is not indicative of a problem.
\end{sloppypar}

\vspace{-1mm}
\subsubsection*{{\bf Rare event emitter}}
If an event emitter is infrequently used, for example because it belongs to a \emph{rarely-used API} or \emph{custom API extension}, then we will not see many listener registrations on this emitter overall.
In particular, any listener-registration pair involving this emitter will appear to be rare (when compared to the entire set of listener-registration pairs collected).

As an example, consider the following listener registration from the GitHub project \texttt{martindale/soundtrack.io}:

\begin{lstlisting}
req.spotify.get(url).on('complete', /* omitted */)
\end{lstlisting}

Here, \code{req} is an instance of \code{http.ClientRequest}. 
Objects of this class do not normally have a \code{spotify} property. 
This is a custom property added by \texttt{soundtrack.io} for interacting with the Spotify API.\footnote{Recall that in JavaScript the properties of an object are not fixed; properties can be added, overwritten, and deleted dynamically.} Consequently, we see the access path of this registration very infrequently; we only encountered it twice in our evaluation.

Further study of the source code reveals that \code{req.spotify.get} does, indeed, return an event emitter that supports the \code{complete} event, so this listener registration is \ValidationSetCorrectTerm{} in this context.

\vspace{-1mm}
\subsubsection*{{\bf Rare event name}}\label{sec:rare-event-name}
The Node.js event API allows client code to emit and register listeners for \emph{custom events}. 
Hence, a listener-registration pair may be infrequent simply because the event is a custom event that is only used in one particular code base.

For example, the test suite of the \code{emitter-listener} npm package~\cite{EmitterListener} uses a custom event \code{test} on \code{http.ServerResponse} objects. 
This is encountered three times in this particular code base, and all three instances correspond to \ValidationSetCorrectTerm{} usages of the custom event.
We would not expect this pair to appear anywhere else (and indeed we did not find any other instances in our evaluation), but in spite of its rarity it is a \ValidationSetCorrectTerm{} pair.

To avoid the above two situations, it is not enough to consider the rarity of the pair when compared to all other pairs.
We want to only classify a listener-registration pair $\Pair{a}{e}$ as \LearnedIncorrectTerm{} (and hence potentially buggy) if \textit{both} of the following hold:
\begin{enumerate}
\item $e$ is a rare event for $a$;
\item $a$ is a rare access path for $e$.
\end{enumerate}

The first condition excludes rare event emitters, as in the example from \texttt{martindale/soundtrack.io} above:
the access path only occurs in two listener-registration pairs, one of which registers the \code{complete} event. 
Hence \code{complete} appears in 50\% of all pairs involving the access path, meaning that it is (intuitively) not a rare event for the access path.

\begin{sloppypar}
The second condition excludes rare event names, as in the example from \texttt{emitter-listener}:
the \code{test} event only occurs in three listener-registration pairs,%
\footnote{Recall that we disambiguate event names based on the root package of the access path we see them registered on. 
While many packages have a \code{test} event, in this case we are only interested in \code{test} events related to the \code{http} package.}
one of which registers it on $\mathbf{require}(\mathtt{http}).\mathtt{ServerResponse}_{\mathbf{new}}()$. 
Hence this access path appears in 33\% of all pairs involving the \code{test} event, meaning that it is (intuitively) not a rare access path for the event.
\end{sloppypar}

We now develop a statistical analysis to make this intuitive notion of rarity rigorous and effectively computable.

\subsection{Statistical analysis}
\newcommand{\BCDF}{\text{BCDF}}

To motivate our statistical analysis, let us first consider the problem of
determining whether an access path $a$ is rare for an event $e$; the converse
problem of determining whether an event $e$ is rare for an access path $a$ is
handled symmetrically.

To determine if an access path $a$ is rare for an event $e$, we would like to measure the probability $p$ that for an arbitrary
listener-registration pair $\Pair{a'}{e}$ we find $a'=a$. If this probability
is smaller than some \emph{rarity threshold} $p_a$, then $a$ is rare for $e$.
For example, $p_a = 0.05$ means that we consider an access path $a$ rare for
an event $e$ if it occurs in less than 5\% of all listener registrations
involving $e$. Since it is not a priori clear what value to choose for
$p_a$, we turn it into a parameter of the statistical analysis. In
Section~\ref{EvalSection} we will empirically evaluate different choices for
$p_a$.

To determine how often $a$ appears in pairs involving $e$, we have to look at all $n_e$ pairs $\Pair{a'}{e}$ and count the number where $a' = a$.
This is modeled as a sequence of $n_e$ experiments, and these experiments are independent, since checking if one pair has an access path matching $a$ has no effect on the checking of any other pairs.
Hence it makes sense to model the probability $p$ of such an $a'$ being the $a$ we are
interested in as a binomial distribution. In general, the binomial
distribution describes the probability distribution of the number of
``successes'' in a sequence of independent experiments, where in this case
success means $a'$ being $a$.

We cannot measure the probability $p$ directly, since our data set
only covers a small fraction of the universe of all
existing or possible JavaScript code. Instead, we use a confidence test to
determine how likely it is, based on our limited data set, that $a$ is a rare
access path for $e$, that is, that the true (but unknown) probability $p$ is
smaller than $p_a$.

As is usual for hypothesis tests, we will actually test the converse: how
\emph{un}likely it is that $a$ is a \emph{common} access path for $e$, that
is, that $p\geq p_a$.

Since we model the probability of an access path occurring with an event as a
binomial distribution, we can use the binomial cumulative distribution
function (BCDF) to implement this test~\cite{bcdfReference}: 
for a series of
$n_e$ independent experiments, $\BCDF(k, n_e, p_a)$ is the likelihood that the
probability of success is at least $p_a$, assuming that at most $k$ of the
experiments are successful.

In our setting, the $n_e$ experiments we consider are checks of all listener
registration pairs of the form $\Pair{a'}{e}$, and $k$ is the number of pairs
where $a'=a$. As explained above, $p_a$ is the rarity threshold we use to
classify an access path as rare for an event.
If this likelihood $\BCDF(k, n_e, p_a)$ is \emph{less} than a (small)
\emph{confidence threshold} $p_{ca}$, then this means that based on our
data it is \emph{unlikely} that $p$ is at least $p_a$, and so it is
\emph{likely} that it is, in fact, less than $p_a$.

As a concrete example, for the pair
$\Pair{\mathbf{require}(\mathtt{http}).\mathtt{request}(1)(0)}{\mathtt{timeout}}$
corresponding to the bug in Figure~\ref{fig:running-example} we have $n_e=216$
and $k=2$: the \code{timeout} event occurs in 216 pairs, but only twice with
this access path. 
Intuitively, since we see this access path in $2/216$ pairs, we might expect a $p$ value around $0.01$, 
but higher values like $p>0.05$ seem unlikely.
\begin{sloppypar}
Plugging in these values into the BCDF formula, we get $\BCDF(2, 216, 0.05) \approx
0.001$, meaning that based on our observations the likelihood of $p$ being
greater than 0.05 is 0.1\%. Turning this statement around, we are 99.9\%
certain that $a$ occurs in 5\% or less of all access pairs involving $e$. Now,
to conclude that $a$ is indeed rare for $e$ (with the rarity threshold $p_a =
0.05$), we need this 99.9\% certainty to satisfy the chosen confidence
threshold $p_{ca}$. If, for example, we chose a $p_{ca} = 0.05$, the
confidence threshold would be 95\% and so we \emph{would} conclude that $a$ is
rare for $e$.
\end{sloppypar}

This confidence threshold $p_{ca}$ is also a parameter of the statistical analysis, so that we
ultimately end up with four parameters: two rarity thresholds $p_a$ and $p_e$,
and two confidence thresholds $p_{ca}$ and $p_{ce}$, all of which range
between 0 and 1 (as they represent probabilities).

The rarity threshold $p_a$ determines when we consider an access path $a$ to
be rare for an event $e$, and the rarity threshold $p_e$ determines when we consider an event $e$ to be rare for an access path $a$. 
The confidence threshold
$p_{ca}$ determines how confident we want to be that $a$ is actually rare
for $e$ based on the data, and similarly for $p_{ce}$.

Putting it all together, then, we consider a listener-registration pair $\Pair{a}{e}$ to be rare if both rarity tests succeed, that is, if the following condition holds:
\begin{equation*}
  \BCDF(k, n_a, p_e) < p_{ce} \; \land \; \BCDF(k, n_e, p_a) < p_{ca}
\end{equation*}

\subsection{Refining the statistical analysis}
Applying this condition in practice, we noticed one particular scenario where it led to misclassifications: if for an event $e$ there are many pairs $\Pair{a}{e}$, but each individual pair occurs infrequently, we will end up classifying all access paths $a$ for this event as rare. 
This pattern arises, for instance, with custom events used in tests.

As a concrete example, there are 522 $\Pair{a}{e}$ pairs registering a listener for the \code{doge} event on an $a$ rooted at the npm package \code{socket.io-client}. 
This nonsensical event name is commonly used for a placeholder or test event -- this is reflected in the data, as we see that these 522 pairs involve 520 unique paths, 519 of which occur in exactly one pair. 
In other words, the usage of \code{doge} follows no discernible pattern in the data.

For one of the pairs $\Pair{a}{\mathtt{doge}}$ where $a$ only occurs once and a rarity threshold $p_a$ of 0.01, we get $\BCDF(1, 522, 0.01) \approx 0.03$, so we would conclude with 97\% confidence that $a$ is rare for \code{doge} and might then label it as \LearnedIncorrectTerm{}. 
This is undesirable: we should not conclude anything about this pair, since the data is too sparse.

We encode this into the statistical analysis by changing the occurrence count $k$ to not only count occurrences of the pair $\Pair{a}{e}$, but also occurrences of pairs $\Pair{a'}{e}$ where $a'$ appears together with $e$ as often or less often than $a$.

Formally, we write $k_e(a)$ for the number of times the pair $\Pair{a}{e}$ occurs in the data (for which we used $k$ above), and then define
\[
k_e(\lceil a\rceil) = \sum\{k_e(a')\mid 0 < k_e(a')\leq k_e(a)\}
\]

Intuitively, this means that we are now not only taking into account the absolute number of times we see $a$ together with $e$, but also how that number compares to that of other $a$s (on the same $e$). 
For example, for the 519 access paths that only appear once together with \code{doge}, we now have $k_e(\lceil a\rceil)=519$, making them very unlikely to be considered rare.

Defining $k_a(\lceil e\rceil)$ symmetrically as the number of occurrences of pairs $\Pair{a}{e'}$ where $e'$ appears together with $a$ as often or less often than $e$, we refine the overall condition for a pair $\Pair{a}{e}$ being classified as \LearnedIncorrectTerm{} as follows:
\begin{equation*}
  \BCDF(k_a(\lceil e\rceil), n_a, p_e) < p_{ce} \; \land \; \BCDF(k_e(\lceil a\rceil), n_e, p_a) < p_{ca}
\end{equation*}

\begin{sloppypar}
  In particular, the single-occurrence access paths above now fail the second condition since $\BCDF(k_e(\lceil a\rceil), 522, 0.01) = \BCDF(519, 522, 0.01) \approx 1$, that is, we are almost 100\% confident that these access paths do not meet the rarity threshold of 0.01.
\end{sloppypar}

It should be noted, however, that this formulation does result in more false negatives: if any of these access paths is actually \ValidationSetIncorrectTerm{}, they will no longer be flagged. 
Since we are mostly interested in automated bug detection, we are willing to trade false positives for false negatives.

\section{Implementation}\label{ImplementationSection}

This section provides some details on the implementation of the two stages of our approach.

For the mining stage, we implemented a static analysis in QL~\cite{DBLP:conf/ecoop/AvgustinovMJS16} for identifying event registrations in JavaScript. Extensive libraries for writing static analyzers in QL are available as part of CodeQL~\cite{qlrepo}, including, in particular, an implementation of access paths, making it an ideal tool for our purposes.
The CodeQL access-path library already performs elimination of cyclic access paths as described in Section~\ref{sec:APAliasRemoval}; we additionally implemented replacement of chained listener registrations.

By writing the mining analysis in QL we were moreover able to leverage LGTM.com~\cite{lgtm.com}, a cloud-based analysis platform that, at the time of writing, makes over 130,000 open-source code bases available for analysis. Out of these, around \TotalProjectsApprox{} contain at least some JavaScript code, which we use as the basis of the evaluation in Section~\ref{EvalSection}.

The statistical analysis detailed in Section \ref{ClassificationSection} is implemented in Python, using the pandas library~\cite{pandas}, and the SciPy library~\cite{scipy} for the statistical computations.

\section{Experimental Methodology}\label{MethodologySection}

This section covers a number of points related to the setup and methodology used in the experimental evaluation that we will
report on in the next section.  

To evaluate the statistical analysis that was presented in Section~\ref{ClassificationSection}, we need to compare
its inferred classifications against a ``ground truth'' set of pairs with known classifications. Here, the challenge
is that the sheer size of the data set precludes exhaustive manual validation. 
Therefore, we adopted an approach in which a \textit{validation set} is constructed semi-automatically for those 18 packages for 
which the largest number of event registrations were found during the mining phase%
\footnote{
  {\tt http}, {\tt net}, {\tt fs}, {\tt process}, {\tt child\_process}, {\tt https}, {\tt socket.io}, {\tt socket.io-client}, {\tt stream}, {\tt readable-stream}, {\tt events}, {\tt cluster}, {\tt zlib}, {\tt ws}, {\tt readline}, {\tt http2}, {\tt repl}, {\tt tls}
}, based on a manual analysis of the API documentation. 
Section~\ref{sec:ConstructingValidationSet} reviews this approach. 
Then, Section~\ref{sec:MeasuringAnalysisQuality} defines how the usual notions of {\it recall} and {\it precision} can be defined by
correlating the results of the statistical analysis with the validation set.    
Finally, Section~\ref{sec:SetupStatisticalAnalysis} provides details on the configuration and thresholds used for the data mining and 
statistical analysis.

\subsection{Constructing a Validation Set} \label{sec:ConstructingValidationSet}

We construct a validation set of $\Pair{\texttt{access path}}{\texttt{event}}$ pairs that are labeled in one of three ways:
\begin{itemize}
  \item
    \textit{\ValidationSetCorrectTerm{}} pairs reflect API usage that is correct to the best of our knowledge,
  \item
    \textit{\ValidationSetIncorrectTerm{}} pairs reflect API usage that is incorrect to the best of our knowledge, and
  \item
    \textit{\ValidationSetImpreciseTerm{}} pairs contain an imprecise access path,
    meaning that the same pair can correspond to both correct and incorrect usages (see Section~\ref{ImpreciseAccessPathsSection}).
\end{itemize}
To allow a meaningful comparison, the validation set only contains pairs that are also present in the data set produced by the mining stage.
Since the labeling process relies on manual interpretation of API documentation, it is subject to human error as discussed in Section~\ref{CaseStudyAndThreatsSection}.

We first consider how \textit{\ValidationSetCorrectTerm{}} pairs in the validation set are identified.
For each of the packages under consideration, we studied the documentation and made lists of the events emitted and access paths exported by each API.
This manual analysis typically involves observing that a specific API call returns an object (or receives an object as an argument)
that emits a specific set of events. 
For example, the documentation for the \code{http} package\footnote{
 See \url{https://nodejs.org/api/http.html}.
} states that the \code{request} function exported by this package 
  ``returns an instance of the \code{http.ClientRequest} class. The \code{ClientRequest} instance is a writable stream.'', and that writable-stream objects emit the \code{error} event.
From this we can conclude that objects represented by the access path $\mathbf{require}(\mathtt{http}).\mathtt{request}()$
emit the \code{error} event, so the pair $\Pair{\mathbf{require}(\mathtt{http}).\mathtt{request}()}{\code{error}}$ reflects a 
\textit{\ValidationSetCorrectTerm{}} use of the \code{http} API.
 
Furthermore, the documentation for \code{http} states that, in a successful HTTP request,  
events \code{data} and \code{end} may be emitted on the object that is passed as
the first argument to the callback that is passed as the second argument to \code{request}. From this, we conclude
that the pairs
  $\Pair{\mathbf{require}(\mathtt{http}).\mathtt{request}(1)(0)}{\code{data}}$
and
  $\Pair{\mathbf{require}(\mathtt{http}).\mathtt{request}(1)(0)}{\code{end}}$
also reflect \textit{\ValidationSetCorrectTerm{}} usage of the \code{http} package.

Next, we consider how a set of \textit{\ValidationSetIncorrectTerm{}} pairs for a given API is added to the 
validation set. To this end, we simply construct the set of all events that are \textit{not} emitted by the API itself, but
that are emitted by some other API in the same package.
For example, from examining the {\tt http} API documentation, it can be seen that a call to {\tt http.createServer} 
returns an object that emits {\tt connection} events. Since this event is not documented as being emitted by the
object that is returned by \code{http.request}, the validation set contains a pair
  $\Pair{\mathbf{require}(\mathtt{http}).\mathtt{request}()}{\code{connection}}$ that reflects an
\textit{\ValidationSetIncorrectTerm{}} use of the API. 

Finally, the set of \textit{\ValidationSetImpreciseTerm{}} pairs is constructed by taking all pairs
whose access path references the second callback parameter of the \texttt{on} method as well as a few other methods used
for event registration: as explained in Section~\ref{ImpreciseAccessPathsSection}, these access paths do not include
information about which event the listener is being registered for, so the access path may correspond to both correct
and incorrect API usages.
Taking the examples from Section \ref{ImpreciseAccessPathsSection}, the pairs $\Pair{\mathbf{require}(\mathtt{http}).\mathtt{request}().\mathtt{on}(1)(0)}{\code{response}}$ and $\Pair{\mathbf{require}(\mathtt{http}).\mathtt{request}().\mathtt{on}(1)(0)}{\code{socket}}$ are labeled as imprecise.
For similar reasons we also mark access paths involving the built-in JavaScript methods
\texttt{apply}, \texttt{bind} and \texttt{call}, which perform reflective operations on functions, as imprecise.
The entire generated model for {\tt http} can be found in the supplemental materials associated with this paper%
\footnote{
  Moreover, all generated models, and the model-generating script is available at: \href{https://github.com/emarteca/JSEventAPIModelling}{\textcolor{blue}{https://github.com/emarteca/JSEventAPIModelling}}.
}.
 
As mentioned above, all pairs in the validation set also appear in the mined data set. However, the converse is not true,
mainly for two reasons\footnote{
Supplemental materials provide some examples of pairs in our mined data that are not included in the validation set.
}:
\begin{enumerate}
  \item
    The access path may be rooted in the import of an API that is not among the 18 packages considered in the validation set.
  \item
    The access path and/or the event are not mentioned in the API documentation. This is, for example, the case for custom
    API extensions or events like the \code{spotify} property in the example in Section~\ref{ClassificationSection},
    or for deprecated APIs that have been removed from the documentation.
\end{enumerate}

\subsection{Measuring analysis quality} \label{sec:MeasuringAnalysisQuality}
Given the validation set, we can now define the usual metrics for assessing the effectiveness of the statistical analysis.

A {\it false positive} is a pair $\Pair{\texttt{access path}}{\texttt{event}}$ that is classified as \LearnedIncorrectTerm{} by the
statistical analysis but that is labeled as \ValidationSetCorrectTerm{} or \ValidationSetImpreciseTerm{}\footnote{
We consider pairs that the statistical analysis classifies as being \LearnedIncorrectTerm{} but that are labeled as imprecise in the validation set as false positives, to report the most pessimistic results for our technique.
} in the validation set.
Conversely, a {\it false negative} is a pair that is labeled as \ValidationSetIncorrectTerm{} in the validation set but that
the statistical analysis does not classify as \LearnedIncorrectTerm{} (i.e., we consider both the case where the statistical analysis
classifies it as \LearnedCorrectTerm{} and the case where the statistical analysis has left it unclassified as false negatives).
Finally, a {\it true positive} is a pair that is labeled as \ValidationSetIncorrectTerm{} in the validation set and that the
statistical analysis also classifies as \LearnedIncorrectTerm{}.

In Section \ref{CaseStudySection} we investigate reasons for false positives and false negatives that occur in the results of the statistical analysis,
and present a case study of some specific demonstrative examples.

Based on these definitions, we now can now define the {\it recall} of the statistical analysis as the percentage of pairs labeled as \ValidationSetIncorrectTerm{}
in the validation set that are classified as \LearnedIncorrectTerm{} by the statistical analysis.
Moreover, the {\it precision} of the statistical analysis is defined as the percentage of true positives among all \LearnedIncorrectTerm{} pairs reported by the statistical 
analysis.

\subsection{Data Mining and Statistical Analysis} \label{sec:SetupStatisticalAnalysis}

We ran the mining analysis on all \TotalProjects{} JavaScript projects available on LGTM.com at the time of writing this paper. 
With this, we collected a total of \TotalPairs{} $\Pair{a}{e}$ listener-registration pairs (\TotalUniquePairs{} of which were unique), 
from \ProjectsUsingEvents{} projects. The remaining projects did not use event-based APIs recognized by the analysis.

Of this mined data, we have labeled \TotalLabeledCorrect{} pairs as being \ValidationSetCorrectTerm{} API uses, \TotalLabeledImprecise{} pairs as having an imprecise access path, and \TotalLabeledIncorrect{} as \ValidationSetIncorrectTerm{} API uses, for a total of \TotalLabeledPairs{} labeled pairs forming our validation set. 

Section~\ref{EvalSection} will explore how recall and precision of the statistical analysis are affected by the selection
of different parameters for the rarity thresholds $p_a$ and $p_e$ and the confidence thresholds $p_{ca}$ and $p_{ce}$, and
which configurations generally yield the best tradeoff between recall and precision.
For these experiments, we needed to choose a set of parameter values to test.
For the rarity thresholds $p_a$ and $p_e$, we chose values from the set $\{0.005, 0.01, 0.02, 0.03, 0.04, 0.05, 0.1, 0.25\}$.
A value of $p_a=0.005$, for instance, means that we consider an access path to be rare for an event if it occurs in less than 0.5\% of all pairs with this event.
For the confidence thresholds we chose values from the set $\{0.005, 0.01, 0.02, 0.03, 0.04, 0.05, 0.1, 1\}$.
A value of $p_{ca}=0.005$, for instance, means that we want to be 99.5\% sure that an $a$ is rare for an $e$ before classifying it as rare.
The extreme value of $p_a=1$ has the effect of classifying every $a$ as rare for $e$, thereby reducing the statistical analysis to just checking whether events are rare for access paths (and vice versa).
This allows us to test the sensitivity of the statistical analysis to the rarity of access paths and events individually.

Altogether, this results in a space of 4,096 configurations.

\section{Evaluation}\label{EvalSection}

To evaluate the practicality of our approach, 
we run the statistical analysis over the mined data for each of the 4,096 configurations, and 
assess the results quantitatively and qualitatively with the following research questions:

\begin{itemize}
\item[\textbf{RQ1.}] {\it Impact of configuration parameters on precision/recall}: How do precision/recall change as the configuration parameters vary?
\item[\textbf{RQ2.}] {\it Impact of training set selection on precision/recall}: How do precision/recall change as the training set selection varies?
\item[\textbf{RQ3.}] {\it Impact of training set size on precision/recall}: How do precision/recall change as the training set size varies?
\item[\textbf{RQ4.}] {\it Utility of results}: Does the approach identify practically relevant mistakes?
\item[\textbf{RQ5.}] {\it Performance}: Is the approach practical in terms of performance/resources?
\end{itemize}

\vspace{10pt}
\noindent
We now address each of these research questions in turn.

\subsection{RQ1: Impact of configuration parameters on precision/recall}

Since the goal is to automatically find dead listener patterns, our approach has to achieve two things to be practically useful: 
it should classify as \LearnedIncorrectTerm{} as many as possible listener-registration pairs  that are labeled as \ValidationSetIncorrectTerm{} in the validation set, 
while at the same time minimizing the number of pairs that are classified as \LearnedIncorrectTerm{} that are labeled as \ValidationSetCorrectTerm{} in the validation set.
In other words, we should maximize for both recall \emph{and} precision.

How well the approach achieves these goals depends on the parameters of the statistical analysis, so we systematically explore the space of parameter configurations to find one that maximizes recall while maintaining an acceptable precision rate (defined as 90\% in accordance with the literature~\cite{Sadowski15,Sadowski18}).
For each configuration we run the classification to find \LearnedIncorrectTerm{} listener-registration pairs,
and use the validation set described in Section \ref{sec:ConstructingValidationSet} to determine the precision and recall of the statistical analysis.

\begin{figure*}
\centering
\includegraphics[width=0.75\textwidth]{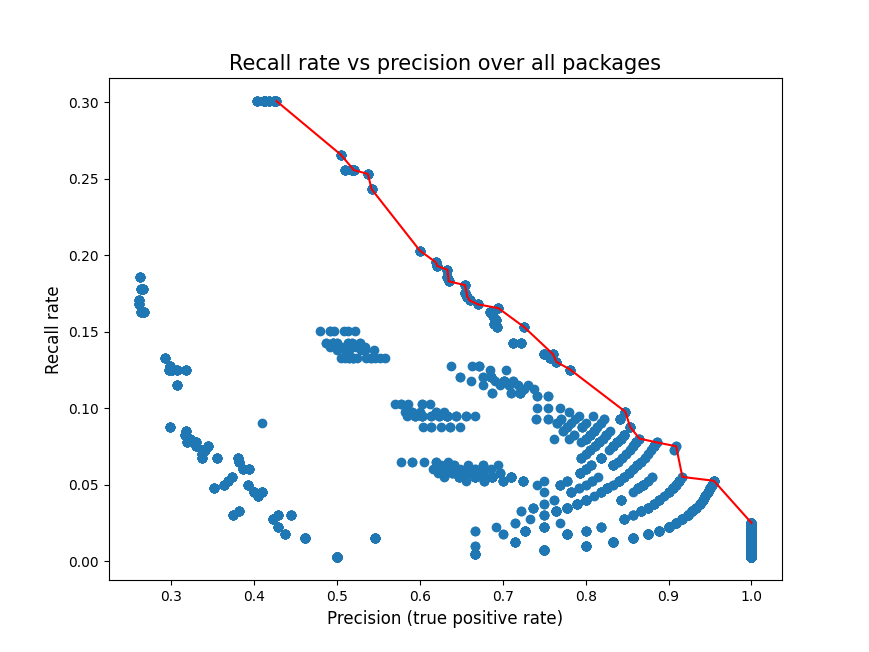}
\caption{Precision and recall for all configurations (blue dots); Pareto front in red}\label{ParetoGraph}
\end{figure*}

Figure~\ref{ParetoGraph} shows the results of this experiment. 
Unsurprisingly, there is an inverse correlation between the recall and precision: configurations that classify many pairs as being \LearnedIncorrectTerm{} have many true positives, but also many false positives. 
Hence it is not meaningful to optimize either metric in isolation.

Instead, we want to concentrate on the \emph{Pareto front}~\cite[Chapter~16]{mas1995microeconomic}, that is, the set of configurations for which there is no other solution that is better on both metrics (the red line in Figure~\ref{ParetoGraph}): a configuration is in the Pareto front if there is no configuration with the same (or higher) precision that has a higher recall.

Altogether, there are eight configurations on the Pareto front with precision of 80\% or above, as detailed in Table~\ref{fig:BestConfigurations}.
For each configuration we show the values of the four parameters, the precision and recall, and the number of unique true positives,  false positives, and pairs in the validation set that remain unclassified by the statistical analysis.
We also show the number of times these true positives occur in the entire data set (roughly speaking, this is the number of potential bugs the configuration finds) and the number of projects they occur in.
For example, the first row reads as follows: a parameter configuration of $p_a,\, p_e,\, p_{ca},$ and $p_{ce}$ as 5\%, 5\%, 2\%, and 10\% respectively, results in a precision of 100\% and recall of 3.0\% over the validation set.
This corresponds to 12 true positives (\#TP), no false positives (\#FP), and no unclassified pairs (\#UP); these true positive pairs occur 23 times in the mined data (Occ TP), across 22 projects (\#Proj).

\begin{table*}
\centering
{\small
  \begin{tabular}{ l || r | r || r | r | r || r | r } 
  \multicolumn{1}{l||}{{\bf Configuration}} & \multicolumn{7}{c}{{\bf Results}}\\
  \hline 
  $(p_a, p_e, p_{ca}, p_{ce})$ & \% Precision & \% Recall & \# TP & \# FP & \# UP & Occ TP & \# Proj \\
  \hline\hline
  (0.05, 0.05, 0.02, 0.1) & 100.0 & 3.0 & 12 & 0 & 0 & 23 & 22 \\
  (0.1, 0.05, 0.05, 0.1) & 95.8 & 5.8 & 23 & 1 & 0 & 57 & 48 \\
  (0.1, 0.05, 0.1, 0.1) & 92.3 & 6.0 & 24 & 2 & 0 & 58 & 49 \\
  \rowcolor{lightgray} (0.1, 0.1, 0.03, 0.01) & \BestTPRate{} & \BestRecallRate{} & 30 & 3  & 1 & \TotalIncorrectPairs{} & 64 \\
  (0.25, 0.04, 0.01, 0.005) & 88.6 & 7.8 & 31 & 4 & 3 & 77 & 61 \\
  (0.25, 0.05, 0.01, 0.01) & 86.5 & 8.0 & 32 & 5 & 3 & 79 & 63 \\
  (0.25, 0.01, 1, 0.04) & 85.4 & 8.8 & 35 & 6 & 9 & 48 & 36 \\
  (0.25, 0.01, 1, 0.1) & 84.8 & 9.8 & 39 & 7 & 12 & 55 & 41 \\
  \end{tabular}
}
  \caption{Configurations with $\geq$80\% recall; optimal configuration highlighted in gray.}\label{fig:BestConfigurations}
\end{table*}

To answer RQ1, then, we found that there are indeed configurations with more than 90\% precision.
The fourth row represents the configuration we consider optimal:
This is the configuration that yields the highest recall for a precision over 90\%.
The rarity thresholds $p_a$ and $p_e$ are 10\% and 10\%, and the confidence thresholds $p_{ca}$ and $p_{ce}$ are 3\% and 1\%, respectively.
Over the validation set, this configuration yields three false positives and 30 true positives, for a precision of \BestTPRate{}\%.
The true-positive pairs occur \TotalIncorrectPairs{} times in total across 64 projects.
All the false positives for this configuration were cases where the access path is overly imprecise.

\subsection{RQ2: Impact of training set selection on precision/recall}
The configuration we identified as optimal in RQ1 performs very well on the full validation set, but of course this does not imply that it would do as well on another data set.
In order to address this concern without having to manually label even more pairs, we conducted a 10-fold cross-validation experiment.
We divided the data into 10 random partitions. Then, we determine the best configuration (i.e., the highest recall with at least 90\% precision) over nine of these partitions (the training data) and validate them on the remaining partition (the validation data). We repeat this procedure ten times, once for each of the partitions as the validation data.

\begin{table*} 
\centering
{\small
  \begin{tabular}{ c | l | r | r | r | r | r | r } 
  \textbf{Round} & \textbf{Configuration} & \multicolumn{3}{c|}{{\bf Training}} & \multicolumn{2}{c}{{\bf Validation}} \\
  \hline 
  & $(p_a, p_e, p_{ca}, p_{ce})$ & \% Precision & \% Recall & \# TP & \% Precision & \% Recall & \# TP \\
  \hline\hline
  0 & (0.25, 0.04, 0.01, 0.005) & 90.6 & 8.1 & 29 & 87.5 & 5.0 & 2 \\
  1 & (0.25, 0.04, 0.01, 0.005) & 91.2 & 8.6 & 31 & 75.0 & 7.5 & 3 \\
  2 & (0.1, 0.1, 0.03, 0.01) & 92.0 & 6.4 & 23 & 87.5 & 17.5 & 7 \\
  3 & (0.1, 0.05, 0.1, 0.1) & 90.9 & 5.6 & 20 & 100.0 & 10.0 & 4 \\
  4 & (0.1, 0.05, 0.1, 0.1) & 91.7 & 6.1 & 22 & 100.0 & 5.0 & 2 \\
  5 & (0.1, 0.1, 0.04, 0.01) & 90.3 & 7.8 & 28 & 75.0 & 5.0 & 2 \\
  6 & (0.25, 0.04, 0.01, 0.005)  & 90.0 & 7.5 & 27 & 80.0 & 10.0 & 4 \\
  7 & (0.1, 0.1, 0.03, 0.02) & 90.6 & 8.1 & 29 & 87.5 & 5.0 & 2 \\
  8 & (0.1, 0.1, 0.03, 0.01) & 90.3 & 7.8 & 28 & 100.0 & 5.0 & 2 \\
  9 & (0.1, 0.1, 0.03, 0.02) & 90.3 & 7.8 & 28 & 100.0 & 7.5 & 3 \\
  \end{tabular}
}
  \caption{Outcomes of cross-validation experiment}\label{fig:cross-validation}
\end{table*}

The results of the experiment are shown in Table~\ref{fig:cross-validation}.
Each row represents the results of using a group of nine partitions as the training data and the remaining partition as the validation data.
The second column shows the optimal configuration over the training data.
Columns 3-5 show the precision, recall, and absolute true positive count on the training data, while columns 6-8 show the same on the validation set.
For example, the first row reads as follows: in the first round the optimal configuration on the training data was $p_a=0.25$, $p_e=0.04$, $p_{ca}=0.01$, $p_{ce}=0.005$, which achieved a 90.6\% precision with 8.1\% recall, finding 29 true positive results.
On the validation data, that same configuration resulted in a precision of 87.5\% and recall of 5.0\%, with 2 true positive results.

We see consistent results with the cross-validation experiment.
Concretely: across the 10 rounds of the experiment, in the training data we see an average precision of 90.8\% (standard deviation 0.7\%) and an average recall of 7.4\% (standard deviation 1.0\%).
Then, in the validation data we see an average precision of 86.3\% (standard deviation 10.3\%) and an average recall of 7.8\% (standard deviation 4.0\%).
From this we see that not only is the quality of results consistent between training runs, but that it also results in consistent results on the validation data.

Looking at the configurations determined to be optimal, we see a high occurrence rate of each of the parameters determined optimal over the whole set.
Concretely: the optimal $p_a = 0.1$ is found in 7 runs and $p_a$ = 0.25 (resulting in a precision of 88.6\% over the whole set) is found in the other 3.
Similarly, the optimal $p_e = 0.1$ is found in 5 runs, $p_{ca} = 0.03$ in 4 runs, and $p_{ce} = 0.01$ in 3 runs.
In conclusion, the choice of training data does not substantially affect the choice of optimal configuration.

\subsection{RQ3: Impact of training set size on precision/recall}

Having shown that the \emph{selection} of the training data does not matter much, we will now explore the effect of the \emph{size} of the training data: how well does the statistical analysis perform when trained over smaller data sets?

In order to test this, we designed an experiment where we randomly sampled a given percentage of the data, and then determined the optimal configuration on this subset%
\footnote{
  Alternatively, one could apply the analysis to a subset of projects and examine the stability of the results.
  However, we decided against this approach because the number of modeled API calls varies considerably
  between projects. 
}. As before, we define ``optimal'' to mean the configuration with the highest recall that achieves a precision of at least $90\%$.
Then, we take this configuration and report the precision/recall it achieves over the whole data set.
We repeat this process 10 times for each percentage, and test this on samples of 2\%, 5\%, 10\%, 25\%, and 50\% of the total data.
For each of these sample percentage, we report the average (harmonic mean) precision and recall computed over all 10 iterations.

\begin{table*}
\centering
{\small
  \begin{tabular}{ c | l || r | r || r | r } 
  \multicolumn{6}{c}{{\bf 2\% of data sampled for subset}} \\
  \multicolumn{2}{c||}{} & \multicolumn{2}{c||}{{\bf On subset}} & \multicolumn{2}{c}{{\bf On whole set }}\\
  \hline 
   Iter &  Optimal config &  \% Precision &  \% Recall &  \% Precision & \% Recall \\
  \hline\hline
    1 & (0.25, 0.01, 1, 0.1) & 100.0 & 50.0 & 84.8 & 9.8 \\
    2 & (0.25, 0.01, 1, 0.1) & 100.0 & 25.0 & 84.8 & 9.8 \\
    3 & (0.25, 0.02, 1, 0.005) & 100.0 & 71.4 & 78.1 & 12.5 \\
    4 & (0.25, 0.005, 1, 0.1) & 100.0 & 12.5 & 84.2 & 4.0 \\
    5 & (0.25, 0.02, 1, 0.005) & 100.0 & 42.9 & 78.1 & 12.5 \\
    6 & (0.25, 0.005, 1, 0.1) & 100.0 & 46.2 & 84.2 & 4.0 \\
    7 & (0.25, 0.03, 1, 0.005) & 100.0 & 41.7 & 71.8 & 15.3 \\
    8 & (0.25, 0.01, 1, 0.05) & 100.0 & 50.0 & 84.1 & 9.3 \\
    9 & (0.25, 0.005, 1, 0.05) & 100.0 & 14.3 & 93.3 & 3.5 \\
    10  & (0.25, 0.01, 1, 0.02) & 100.0 & 37.5 & 83.8 & 7.8 \\
  \hline
  \multicolumn{4}{r}{\textbf{Average (harmean): }} & 82.4 & 6.9 \\

  \multicolumn{6}{c}{{\bf 5\% of data sampled for subset}} \\
  \hline 
   Iter &  Optimal config &  \% Precision &  \% Recall &  \% Precision & \% Recall \\
  \hline\hline
    1 & (0.25, 0.03, 0.005, 0.05) & 100.0 &  25.0 & 87.5 & 7.0 \\
    2 & (0.25, 0.04, 0.1, 0.005)  & 90.0 & 30.8 & 82.9 & 8.5 \\
    3 & (0.1, 0.1, 0.02, 0.05)  & 100.0 &  21.1 & 81.6 & 7.8 \\
    4 & (0.25, 0.02, 1, 0.03) & 90.0 & 37.5 & 76.4 & 13.8 \\
    5 & (0.04, 0.02, 0.005, 0.05) & 100.0 &  7.1 & 100.0 & 1.3 \\
    6 & (0.25, 0.02, 1, 0.1)  & 100.0 &  13.6 & 81.3 & 6.5 \\
    7 & (0.25, 0.03, 0.05, 0.03)  & 100.0 &  10.0 & 81.1 & 7.5 \\
    8 & (0.25, 0.005, 1, 0.01)  & 100.0 &  29.4 & 90.0 & 4.8 \\
    9 & (0.25, 0.02, 1, 0.05) & 100.0 &  36.4 & 80.6 & 6.3 \\
    10 & (0.25, 0.02, 1, 0.1)  & 100.0 &  22.7 & 81.3 & 6.5 \\
  \hline
  \multicolumn{4}{r}{\textbf{Average (harmean): }} & 83.8 & 4.8 \\

  \multicolumn{6}{c}{{\bf 10\% of data sampled for subset}} \\
  \hline 
   Iter &  Optimal config &  \% Precision &  \% Recall &  \% Precision & \% Recall \\
  \hline\hline
    1 &  (0.1, 0.05, 0.05, 0.005) & 100.0 & 5.1 & 95.2 & 5.0 \\
    2 &  (0.25, 0.005, 1, 0.05) & 100.0 & 29.0 & 93.3 & 3.5 \\
    3 &  (0.1, 0.04, 0.05, 0.01 & 100.0 & 15.8 & 95.0 & 4.8 \\
    4 &  (0.25, 0.02, 0.1, 0.1) & 100.0 & 13.5 & 81.3 & 6.5 \\
    5 &  (0.05, 0.1, 0.01, 0.01 & 92.9 & 33.3 & 90.5 & 4.8 \\
    6 &  (0.25, 0.005, 1, 0.03) & 90.0 & 22.5 & 92.9 & 3.3 \\
    7 &  (0.25, 0.005, 1, 0.03) & 100.0 & 31.3 & 92.9 & 3.3 \\
    8 &  (0.25, 0.005, 1, 0.1) & 100.0 & 28.1 & 84.2 & 4.0 \\
    9 &  (0.25, 0.02, 0.1, 0.1) & 100.0 & 13.9 & 81.3 & 6.5 \\
    10 &   (0.25, 0.03, 0.05, 0.03) & 90.0 & 20.5 & 81.1 & 7.5 \\
  \hline
  \multicolumn{4}{r}{\textbf{Average (harmean): }} & 88.4 & 4.5 \\

  \multicolumn{6}{c}{{\bf 25\% of data sampled for subset}} \\
  \hline 
   Iter &  Optimal config &  \% Precision &  \% Recall &  \% Precision & \% Recall \\
  \hline\hline
    1 & (0.04, 0.05, 0.1, 0.03) & 92.3 & 10.4 & 90.5 & 4.8 \\ 
    2 & (0.1, 0.01, 0.02, 0.05) & 90.0 & 8.7 & 90.0 & 2.3 \\ 
    3 & (0.25, 0.04, 0.1, 0.02) & 92.6 & 21.6 & 81.0 & 8.5 \\ 
    4 & (0.03, 0.03, 0.05, 0.04)  & 100.0 &  6.9 & 100.0 & 2.0 \\ 
    5 & (0.25, 0.005, 1, 0.03)  & 90.0 & 12.0 & 92.9 & 3.3 \\ 
    6 & (0.25, 0.005, 1, 0.04)  & 100.0 &  12.6 & 92.9 & 3.3 \\ 
    7 & (0.25, 0.005, 1, 0.05)  & 100.0 &  15.2 & 93.3 & 3.5 \\ 
    8 & (0.25, 0.03, 0.02, 0.02)  & 91.7 & 11.8 & 84.4 & 6.8 \\ 
    9 & (0.25, 0.005, 1, 0.03)  & 90.9 & 9.8 & 92.9 & 3.3 \\ 
    10&   (0.1, 0.05, 0.05, 0.05) & 90.9 & 8.5 & 95.7 & 5.5 \\ 
  \hline
  \multicolumn{4}{r}{\textbf{Average (harmean): }} & 91.0 & 3.6 \\

  \multicolumn{6}{c}{{\bf 50\% of data sampled for subset}} \\
  \hline 
   Iter &  Optimal config &  \% Precision &  \% Recall &  \% Precision & \% Recall \\
  \hline\hline
    1 & (0.05, 0.05, 0.1, 0.05) & 100.0 & 7.0 & 92.9 & 3.3 \\ 
    2 & (0.05, 0.05, 0.05, 0.1) & 93.8 & 7.9 & 92.9 & 3.3 \\ 
    3 & (0.02, 0.05, 0.1, 0.02) & 100.0 & 5.9 & 100.0 & 1.8 \\ 
    4 & (0.04, 0.05, 0.01, 0.02) & 92.9 & 6.6 & 100.0 & 2.8 \\ 
    5 & (0.25, 0.03, 0.02, 0.04) & 94.4 & 8.1 & 84.8 & 7.0 \\ 
    6 & (0.25, 0.04, 0.01, 0.005) & 90.0 & 13.2 & 88.6 & 7.8 \\ 
    7 & (0.25, 0.01, 1, 0.04) & 92.6 & 12.6 & 85.4 & 8.8 \\ 
    8 & (0.05, 0.05, 0.05, 0.04) & 92.9 & 7.2 & 92.3 & 3.0 \\ 
    9 & (0.25, 0.02, 0.1, 0.005) & 91.3 & 10.2 & 81.5 & 5.5 \\ 
    10 &  (0.25, 0.04, 0.02, 0.005) & 90.5 & 9.6 & 86.1 & 7.8 \\ 
  \hline
  \multicolumn{4}{r}{\textbf{Average (harmean): }} & 91.0 & 3.9 \\
  \end{tabular}
}
  \vspace*{1mm}
  \caption{Optimal configurations over smaller percentages of the data, and the corresponding precision/recall over the whole dataset}\label{fig:training-set-size-table}
\end{table*}

Table \ref{fig:training-set-size-table} presents the results of this experiment, with one row per sampling of a given percentage.
For example, the first row can be read as follows: for the first random sampling of 2\% of the data, the optimal configuration is $(0.25, 0.01, 1, 0.1)$.
This achieves a precision of 100\% and a recall of 50\% over the 2\% subset, and a precision of 84.8\% and a recall of 9.8\% over the whole dataset.

Considering only average precision and recall over the whole dataset for the moment, we can see two interesting trends: when training on 2\% of the data, precision over the whole dataset is fairly low at 82.4\%, and so is recall at 6.9\%. As the amount of training data increases, precision increases as well, reaching 91\% for 25\% of the data. Recall, on the other hand, at first \emph{decreases}, hitting a low of 3.6\% on 25\% of the data, before slightly increasing again to 3.9\% at 50\%. (Compare this to the 7.5\% recall reported in RQ1 when training on all 100\% of the data.) This suggests that stable precision can already be achieved with relatively little training data, but more data is needed to improve recall.

Looking more closely at the individual configurations determined to be optimal, we see that for 2\% these configurations always have $p_{ca} = 1$, meaning that \emph{all} access paths are considered rare, and the classification is entirely based on $p_e$ and $p_{ce}$. As discussed in Section~\ref{ClassificationSection}, this means that the analysis will not handle custom events very well, and end up spuriously classifying custom events on common access paths as \LearnedIncorrectTerm{}. This, it turn, leads to very unstable precision, where 100\% precision on the training data dwindles to an average of only 82.4\% over the whole data set.
At higher percentages, configurations with $p_{ca} = 1$ become increasingly rare, with only one left at 50\%: as more data becomes available, the approach learns to deal with custom events better, and precision becomes more stable.

Finally, note that recall on the subset (third column) varies wildly between different samplings at smaller percentages. This is because there are few labeled pairs in those datasets (i.e., the validation data is small), meaning that missing one or two pairs will significantly affect recall.

In conclusion, with small training sets the statistical analysis converges to configurations that do not account for custom events, leading to unstable precision and low recall. As more data is provided, precision becomes better and more stable, while recall is still far below what we can achieve by training on 100\% of the dataset.

\subsection{RQ4: Utility of results}

To qualitatively assess the usefulness of the approach, we conducted a study involving finding bugs in open-source projects. 
In this experiment, 
we examined 100 occurrences of listener-registration pairs that were classified as \LearnedIncorrectTerm{} by the statistical analysis, and manually examined the code from which the pair originated. 

How easy it is to determine whether or not the listener registration is actually a bug depends on the complexity of the access path. 
If the access path is complex and involves interprocedural flow, then it may be hard to determine what events are emitted by the object on which the listener is being registered.
In general, however, we can leverage the fact that we know in advance what the possible reasons for false positives are.
As discussed in Section \ref{AnalysisImprecisionSection}, the lack of flow sensitivity in the mining analysis means that we may report access paths as having listeners registered on them that are avoided in the program via explicit reflection on the type of the access path.
Similarly, the lack of context sensitivity means we do not account for explicit emission of events. 
When looking for false positives, developers could look for explicit emission of the event in question, or for cases where there is an explicit check of the type of the access path before the listener is registered. 

For very long access paths (which we define as 
having at least five property reads, method calls, and/or functions parameters separating it from the initial package import), 
then manual analysis to determine if the pair was indicative of a bug would be very time intensive; also, explaining this error to the developers would have been complicated. 
For expediency, in this experiment we decided to exclude such cases, which accounted for 47 of the code snippets.

For the remaining 53 code snippets, the results can be summarized as follows:
\begin{itemize}
  \item 37 were identified as bugs in the code.
  7 of these were on dead or archived repos (where we define ``dead'' as having not been contributed to since 2012); the other 30 we reported.
  \item 5 had listeners registered on a function parameter, which sometimes takes an object that emits the event and sometimes does not. 
  We decided not to report these cases since they only manifest as a bug in some program executions, and it would be difficult to determine and 
  explain to the developers in which particular executions a bug could happen
  \item 6 were false positives because the listener registration code was only executed if the access path was an instance of a class that emits the relevant event (like the example in Figure \ref{fig:typeReflectionExample})
  \item 5 were false positives because the event was explicitly emitted (like the example in Figure \ref{fig:manualEmitExample}) 
\end{itemize}
Tables 3 and 4 in the supplementary materials include details for each of the individual code snippets that we examined.

For those results that appeared to represent real bugs in the code in active projects, we submitted issues to report them to the developers.
Altogether, we reported \NumReportedIssues{} issues across \NumAnalyzedProjects{} different GitHub projects, \NumConfirmedBugs{} of which have been confirmed by the developers as bugs,\footnote{
Two have been addressed, \cite{erisBugLink} and \cite{harakaBugLink}.
} one where the developers did not remember what the code was supposed to do, and 2 that were false positives (due to explicitly emitting the event elsewhere in the project). We did not yet receive a response for the remaining 20 issues.
Links to all reported issues are included in the supplemental materials.

\begin{figure}
{\small
\begin{lstlisting}
const HTTPS = require("https");

request(method, url, auth, body, file, 
        _route, short) {
  const req = HTTPS.request( /* ... */ )
  req.once("abort", () => { /* ... */ }
  ).once("aborted", () => {/* ... */ }
  );
  req.once("response", (resp) => { /* ... */ }); /*#\label{line:resp}#*/
}
\end{lstlisting}
}
\vspace*{-5mm}
\caption{Condensed version of error in {\tt abalabahaha/eris}}\label{fig:githubIssueExample}
\end{figure}
Figure~\ref{fig:githubIssueExample} shows a simplified version of one of the acknowledged bugs from the project \href{https://github.com/abalabahaha/eris}{\code{abalabahaha/eris}}, a Node.js wrapper for interfacing with Discord.
Here we see the {\tt req} variable created from a call to {\tt http.request}, which returns an instance of {\tt http.ClientRequest}.
However, by examining the {\tt http} API documentation, we see that {\tt aborted} is an event emitted by {\tt http\-.IncomingMessage}, and not {\tt http\-.ClientRequest}.
The developers confirmed this as a bug and fixed it by registering the listener on \code{resp} instead (line~\ref{line:resp}), which had been the original intention.

The full list of bugs reported can be found in the supplemental materials.

\subsection{RQ5: Performance}
Here, we discuss performance and resource requirements.

\paragraph*{{\it Data mining and classification}}
Our approach involves mining and classifying listener registration pairs from a large number of projects. The data mining step requires about 404 hours of compute time
for the \TotalProjects{} projects in the data set. Since LGTM.com runs queries concurrently, this
step was completed in about two days. The classification stage is much faster: classifying the pairs for a given configuration takes only 35 to 40 seconds on commodity hardware. 
We expect these steps to be applied infrequently as event-driven APIs tend to evolve slowly,
and our experimental results suggest that the set of optimal analysis thresholds is fairly stable.

\paragraph*{{\it Per-project costs}}
Once an API model has been constructed, it can be used for a variety of purposes, e.g., in a bug-detection tool
that flags uses of event-driven APIs that are likely to be buggy, or in an IDE plugin for smart completion.
Running the mining analysis on a single JavaScript project is quite fast: for 52\% of all projects in the 
data set, the analysis takes ten seconds or less, with another 45\% taking between ten seconds and a minute. 
There are only 151 projects (0.1\%) for which the analysis takes more than ten minutes. 

We consider these results to be encouraging as, while the upfront cost of constructing an API model is
quite high, our experimental results suggest that the per-project costs are sufficiently low to allow integration
of our approach in a realistic continuous-integration workflow.

\section{Discussion}\label{CaseStudyAndThreatsSection}

This section reports on a case study in which we investigated a few specific examples where the statistical analysis produced 
false positives and false negatives, and considers threats to the validity of our results.

\subsection{Case study of false positives and false negatives}\label{CaseStudySection}

\subsubsection{False negatives} 
False negatives are listener-registration pairs that are labeled as \ValidationSetIncorrectTerm{} in the validation set but that the statistical 
analysis does not classify as \LearnedIncorrectTerm{}. 
Whether or not a given pair $\Pair{a}{e}$ is classified as \LearnedIncorrectTerm{} by the statistical analysis is entirely determined by the frequency
with which $a$, $e$, and the combination $\Pair{a}{e}$ occur in the mined data.
Pairs for which both the access path and the event occur very frequently (but the pair itself is rare) will satisfy the criteria for being classified 
as \LearnedIncorrectTerm{} with more statistical analysis parameter configurations than those \ValidationSetIncorrectTerm{} pairs that appear more rarely. 

As an example, consider the pairs:
\[
   \Pair{\mathbf{require}(\mathtt{net}).\mathtt{createServer}()}{\mathtt{end}}
\] 
and
\[ 
   \Pair{\mathbf{require}(\mathtt{net}).\mathtt{connect}().\mathtt{setNoDelay}()}{\mathtt{secureConnect}},
\] 
which are both labeled as \ValidationSetIncorrectTerm{} in the validation set:
\begin{enumerate}
  \item
    For the first pair, the access path $\mathbf{require}(\mathtt{net}).$ $\mathtt{createServer}()$ 
    occurs 1109 times, the event $\mathtt{end}$ occurs 872 times, and they occur as a pair only twice (in the {\tt net} package). 
    In other words, this rare pair is made up of a very common access path and a very common event; indeed, it meets the thresholds of all of the statistical analysis 
    parameter configurations we tested.  Therefore, this pair is always classified as \LearnedIncorrectTerm{}, i.e., it is a \textit{true positive}.  
  \item
    For the second pair,
    the access path $\mathbf{require}(\mathtt{net}).\mathtt{connect}().\mathtt{setNoDelay}()$ occurs twice, the event $\mathtt{secureConnect}$ occurs 26 times, and 
    they occur only once as a pair.  Since the access path is so rare, this \ValidationSetIncorrectTerm{} pair represents 50\% of the uses of this access path, and 
    therefore it is very unlikely to be classified as \LearnedIncorrectTerm{} by the statistical analysis. 
    Indeed, this pair is not classified as \LearnedIncorrectTerm{} with any of the parameter configurations, unless the rarity of the access path is not considered at all.
    Therefore, it is almost always a \textit{false negative}.
\end{enumerate}
In other words, a false negatives may occur in cases where a given access path or event is used rarely, making it difficult for the statistical analysis to conclude that the
particular event-listener registration pair is rarer still.

\subsubsection{False positives}

The false positives that we observed correspond to event-listener registration pairs that are labeled as \ValidationSetCorrectTerm{} in the validation set
but that show up rarely in the mined data and thus get classified as \LearnedIncorrectTerm{}.

As an example of a false positive, consider the pair $\Pair{\mathbf{require}(\mathtt{process}).\mathtt{stdin}}{\mathtt{drain}}$ which is one of the three
false positives that arises when the analysis is run with the optimal configuration that achieves a precision of \BestTPRate{}\%.
In the mined data, we see the access path $\mathbf{require}(\mathtt{process}).\mathtt{stdin}$ 1,948 times, the event {\tt drain} 234 times, 
but this pair itself only shows up once (in access paths rooted in {\tt process}).   
The vast majority of the {\tt drain} events (209 of the 234) are seen with $\mathbf{require}(\mathtt{process}).\mathtt{stdout}$. 
This is because {\tt drain} is an event on Writable streams \cite{streamDrainEventDocs}, and according to the documentation \cite{processStdinDocs, processStdoutDocs} {\tt process.stdout} is either a {\tt net.Socket} or a Writable stream while {\tt process.stdin} is either a {\tt net.Socket} or a Readable stream. 
Since {\tt net.Socket}s are Duplex streams (i.e., both readable and writable), according to the documentation registering a listener for {\tt drain} on {\tt process.stdin} is a \ValidationSetCorrectTerm{} use of the API. 
However, from the data it seems that although this is a \ValidationSetCorrectTerm{} API usage, this use is rare, and therefore it ends up being classified as \LearnedIncorrectTerm{} by the statistical analysis.

As another example of a false positive, consider the pair 
  $\Pair{\mathbf{require}(\mathtt{zlib}).\mathtt{createGunzip}()}{\mathtt{drain}}$.
In the mined data, the access path shows up 1,649 times, the event 441 times, but the pair itself only once (in access paths rooted in {\tt zlib}).
The most common events we see with this access path are $\mathtt{data}$ (395 times), $\mathtt{end}$ (388 times), $\mathtt{error}$ (364 times), 
and $\mathtt{close}$ (344 times). These are all events emitted by objects of class $\mathtt{stream.Readable}$~\cite{streamReadableDocs}.
This is noteworthy because, according to the documentation, $\mathtt{zlib.createGunzip()}$ returns a $\mathtt{Gunzip}$ object~\cite{zlibGunzipDocs}, and this inherits 
from $\mathtt{stream.Transform}$ which is a Duplex stream (i.e., both readable and writable)~\cite{zlibZipBaseDocs}. The $\mathtt{drain}$ event is an event on
writable streams only. 
So, it seems that although the streams returned by $\mathtt{zlib.createGunzip()}$  are both readable and writable,
 they are almost always used as readable streams, causing the statistical analysis to flag the rare occurrence where this is not the case as
\LearnedIncorrectTerm{}.

\subsection{Threats to Validity}
We are aware of several potential threats to validity.

Our results depend on the set of code bases that have been mined, and this set may not be representative. 
However, we simply used the set of \textit{all} JavaScript projects on LGTM.com that were available at the time of writing this paper, 
which includes many popular open-source projects, and projects added by users of LGTM.com.
These code bases were not specifically selected for this project, and provide 
a reasonable sample of real-world JavaScript code. 

Our measurements of precision and recall are based on a relatively small set of 
listener-registration pairs that we semi-automatically labeled as \ValidationSetCorrectTerm{} or \ValidationSetIncorrectTerm{} 
(\TotalLabeledPairs{} out of \TotalUniquePairs{} unique pairs) and might not generalize beyond this set. 
Exhaustively labeling all pairs was infeasible, so we focused on the most 
popular packages, to ensure that the results are relevant for widely-used APIs. 
Cross-validation showed that the choice of optimal configuration does not crucially depend on the chosen training data set.

The semi-automatic labeling of pairs in the validation set generation involved a review of API documentation by the authors.
Thus, there was potential for human error in this process: if we misread the documentation, some pairs could be mislabeled in the validation set.
In practice, we saw no examples of this in any of the pairs we examined.

The validation set is itself biased in that it contains a relatively small number of pairs labeled as 
\ValidationSetIncorrectTerm{} (\TotalLabeledIncorrect{} out of \TotalLabeledPairs{}). This affects the accuracy of the reported 
precision since we are much more likely to find that a pair classified as \LearnedIncorrectTerm{} by the statistical analysis is actually \ValidationSetCorrectTerm{} 
(and hence a false positive) than \ValidationSetIncorrectTerm{} (and hence a true positive). 
Consequently, our reported precision \textit{underestimates} the actual precision. 
Remedying this imbalance would only improve the precision. 

There is potential for bias in the generalization of our results from the 18 packages we model to other {\tt npm} packages.
In particular, the results might be affected by the number of events and emitters available in a package.
However, the 18 packages in the validation set cover a range of different scenarios 
in terms of the number of events, emitters, and listener-registration pairs $\Pair{a}{e}$ that constitute \ValidationSetCorrectTerm{}  usage of the API.
These values range from 102 \ValidationSetCorrectTerm{} usages over 12 access paths and 13 events with the {\tt stream} package, to 77 \ValidationSetCorrectTerm{} usages over 15 access paths and 24 events with the {\tt fs} package, to only 2 \ValidationSetCorrectTerm{} usages over one access path and 2 events with the {\tt http2} package.   
Table 7 in the supplemental materials shows these values for all the packages in the validation set.   
Thus, we have some confidence that the approach generalizes over packages with differing numbers of events and emitter objects.

The values chosen for the parameters of the statistical analysis obviously greatly influences the quality 
of results. However, our evaluation considered a large number of different combinations, over which we 
determined the optimal configuration for a particular set of conditions (here, for a precision
of at least 90\%). Moreover, a cross-validation experiment revealed the configuration parameters to
be quite stable across subsets of the data.

Finally, the static analysis used in the mining phase is relatively simple and imprecise, e.g., 
due to inherent imprecision of the access path representation. 
Our evaluation accounted for this by considering all pairs involving imprecise access paths to 
be false positives. A more sophisticated analysis using more precise access paths would  also
increase the precision of the statistical analysis.

\section{Learning lost events}\label{LostEventsSection}

At first glance, it seems that we could apply this same learning approach to the dual problem of finding bug patterns in lost events~\cite{DBLP:conf/oopsla/MadsenTL15}, those events which are emitted but never listened for.
We modified our static analysis to identify event emissions instead of event registrations, and reran the data mining to collect information on this dual problem, across the same set of projects.
From this analysis, we mined a total of 22,900 $\Pair{a}{e}$ pairs (10,432 unique).

From this data, we determined that the learning approach cannot be effectively used to identify bug patterns in lost events, since the vast majority of events emitted in projects are \emph{custom events}.
The use of a custom event is specific to the project it appears in, and so patterns observed in other projects cannot be used to learn about its proper use.
We discussed this in Section \ref{sec:rare-event-name} with respect to listener registrations, but the same logic applies to event emissions.

For the remainder of this section, we discuss some details about the data mined on events emitted.

\paragraph*{Base event emitter}
Of the 22,900 pairs mined, we observed that 5,248 have an emitter access path of $\mathbf{require}(\mathtt{events}).\mathtt{EventEmitter}_{\mathtt{new}}()$ and 1,220 have access path $\mathbf{require}(\mathtt{events})_{\mathtt{new}}()$.
No other access paths occur this frequently in our data set.

In code, these access paths correspond to \code{new require('events').EventEmitter()} and \code{new require('events')()} respectively.
Looking at the documentation of the EventEmitter class\footnote{
\href{https://nodejs.org/api/events.html\#events\_class\_eventemitter}{\textcolor{blue}{https://nodejs.org/api/events.html\#events\_class\_eventemitter}}
}, we see that these are aliases, as the EventEmitter class is the default export of the \code{events} package.
Looking at the same documentation, we see that there are only 2 events that make up this API: \code{newListener} and \code{removeListener}.
Any other events emitted on objects that are instances of EventEmitter are therefore custom events. 
With our data, 6,466 of these 6,468 pairs emit a custom event on objects of the base event emitter class.

We examined what users actually do with the custom events on base event emitters.
From our exploration, a few common patterns  of how users build their own custom event infrastructures on top of the base EventEmitter class could be observed; examples of these are included below.

Developers often create custom EventEmitter classes extending the base {\tt EventEmitter} class in a classic object-oriented style.
Consider the following demonstrative example, condensed from the update manager class in \href{https://github.com/2947721120/vscode/blob/master/src/vs/workbench/electron-main/update-manager.ts}{\textcolor{blue}{\texttt{vscode}}}.
{\small
\begin{lstlisting}
export class UpdateManager 
     extends events.EventEmitter {
  // methods that emit and listen for custom events
  private initRaw(): void {
    // ... 
    this.emit('checking-for-update'); 
  }

  public initialize(): void {
    this.on('checking-for-update', /* ... */ );
  }
}

export const Instance = new UpdateManager();
\end{lstlisting}
}
Since the access path representation does not reason about the inheritance hierarchy, the \code{new UpdateManager()} is represented abstractly as $\mathbf{require}(\mathtt{events}).\mathtt{EventEmitter}_{\mathtt{new}}()$.
Other common custom event usage patterns include extending the EventEmitter \code{prototype}
or including an EventEmitter as a class field.
In each of these cases, the developers are encapsulating the base EventEmitter so as to build their own custom event-based infrastructure.

There are a variety of ways developers make use of custom event infrastructures in their code bases.
Building something on top of the base EventEmitter is one of the most common patterns we observed, as we have discussed above; after this, the next most frequently observed emitter was objects of class \code{socket.io} Socket.
Examining the documentation of the \code{socket.io} API about emitting events, we see that there are no standard events.
Therefore, all events emitted on \code{socket.io} client or server based emitters are custom events.
This corresponds to 8,398 of the pairs (1,911 client-side and 6487 server-side).

\paragraph*{Manual analysis of a subset of remaining pairs}
So far, we have determined that 14,866 or the 22,900 pairs mined correspond to emissions of custom events on either the base event emitter or via \code{socket.io}.
This in turn does not mean that the rest of the pairs in our dataset do not correspond to custom events. 

There are 2,668 unique pairs remaining.
Of these, we manually looked at a random sampling of 200, split across all the APIs\footnote{
All pairs manually analyzed are included in a table in the supplemental materials.
}.
In this manual analysis, we found that 82.5\% correspond to custom events.

From this, we conclude that the vast majority of events explicitly emitted correspond to custom events.
As discussed, the use of a custom event is specific to the project in which the custom emitter is defined and so patterns observed in other projects cannot be used to learn about its proper use.
Therefore, our learning approach cannot be effectively used to identify bug patterns in lost events.

\section{Related work}\label{RelatedWorkSection}

A considerable amount of research has focused on detecting and characterizing bugs in JavaScript applications, including
  bug detection tools using
     static analysis \cite{DBLP:conf/sigsoft/BaeCLR14} and
     dynamic analysis \cite{DBLP:conf/icse/PradelSS15,DBLP:journals/pacmpl/AlimadadiZMT18};
  evaluations of the effectiveness of type systems for preventing bugs \cite{DBLP:conf/icse/GaoBB17};
  development of benchmarks \cite{DBLP:conf/icst/GyimesiVSMBF019}; and
  studies of real-world bugs \cite{DBLP:conf/kbse/WangDGGQYW17}.
 
The most closely related work to ours is by Madsen et al.~\cite{DBLP:conf/oopsla/MadsenTL15}. They
describe a static analysis for detecting dead listeners, lost events and other event-handling bugs
based on the notion of an event-based call graph that augments a traditional call graph with edges
corresponding to event-listener registration, event emission, and callback invocation.
Event-handling bugs are detected by looking for patterns in these augmented call graphs.
Unfortunately, their approach does not scale well because their context-sensitive analysis employs
notions of contexts corresponding to the sets of events emitted and listeners registered, which may
be exponential in the size of the program. This exponential behavior appears to manifest itself in practice,
given that, on their largest subject program (which is a mere 390 LOC), one of their analyses incurs a 
running time of 17 seconds, and the other one does not terminate at all.
Our approach targets only dead listeners, and only those cases where the event the listener is meant to
handle is never emitted (excluding cases where it is emitted at a time when the listener is not
registered). This allows us to use a simple and scalable static analysis in our mining
phase, and rely on statistical reasoning over a large data set to offset the noise.

Unfortunately, it is not possible to compare our technique directly against Madsen's, given that their
implementation was a proof-of-concept static analysis for a small subset of ECMAScript 5.  As such, it did 
not support modern JavaScript features such as classes or promises, which are pervasive in the subject 
applications that we analyzed. Furthermore, upon inquiry, we were informed that Madsen's tool is no longer 
available \cite{MadsenPrivateComm:21}. That said, we investigated the bugs reported 
in Madsen's work and found that, of the 12 real-world bugs considered in their work, three are 
dead-listener bugs of the kind that is targeted by our analysis, and our optimal configuration identifies 
all of them. The others concern dead emits, listeners for custom events, or listeners that are dead due to 
the order in which they were added, and are outside the scope of our work.

Our work also stands in a long line of research viewing bugs as ``deviant behavior'': statistical methods are used to infer beliefs or rules that are implicit in the code, and violations of these rules are flagged as likely defects.

Engler et al.~\cite{DBLP:conf/sosp/EnglerCC01} distinguish between ``MUST beliefs'' and ``MAY beliefs''. The former are directly implied by the code, and often boil down to simple data-flow properties: for example, dereferencing a pointer implies the belief that it is not a null pointer, and a subsequent null check of the same pointer is inconsistent with that belief. MAY beliefs, on the other hand, are patterns such as two functions that are often invoked in a particular order, which might reflect an implicit rule (such as the second one freeing a resource allocated by the first one), or might be a coincidence. They use an analysis based on the z statistic to distinguish the two. Our work also aims to infer a MAY belief, but of a more complex kind than those considered by Engler, since the relationship between event emitters and events is many-to-many. Custom events pose an additional challenge that requires more sophisticated statistical methods than the z statistic.

The PR-Miner system~\cite{DBLP:conf/sigsoft/LiZ05} targets a broader class of rules: using frequent itemset mining, it extracts association rules $A\Rightarrow B$, where $A$ and $B$ are sets of program elements such as function calls. Such a rule expresses the observation that functions containing all elements in $A$ also contain all elements in $B$, with a certain level of confidence. Violations of high-confidence rules are then likely to be bugs. Again, the relationship between event emitters and events does not immediately fit this pattern: association rules are ``forall'' rules in the sense that if all elements in $A$ are present then \emph{all} elements in $B$ must be present. By contrast, we are interested in ``exists'' rules in the sense that in any given context a particular type of event emitter emits \emph{one} event from a certain set, but not necessarily all of them.

WN-Miner~\cite{DBLP:conf/tacas/WeimerN05} and PF-Miner~\cite{DBLP:journals/jss/LiuWBH16} focus more narrowly on the problem of inferring temporal specifications, specifically pairs of functions $f$ and $g$ such that $g$ must always be invoked after $f$, usually because it performs some sort of cleanup. Acharya et al.~\cite{DBLP:conf/sigsoft/AcharyaXPX07} generalize this to inferring partial orders between functions. Gruska et al.~\cite{DBLP:conf/issta/GruskaWZ10}, on the other hand, generalize in a different direction and employ association rules of a similar kind as PR-Miner, but where the sets $A$ and $B$ now contain candidate function pairs, thus allowing inference of context-dependent specifications. Murali et al.~\cite{DBLP:conf/sigsoft/MuraliCJ17} apply a Bayesian framework for learning probabilistic API specifications, which is more robust on noisy and heterogeneous data than more lightweight approaches. Although dead-listener detection shares some general principles with temporal-specification mining, the concrete setup is rather different and it is not immediately obvious that their techniques apply to our problem.

Monperrus et al.~\cite{DBLP:journals/tosem/MonperrusM13} propose \emph{type usages} as a particularly useful kind of specification to infer for object-oriented programs: a type usage is a set of methods invoked on a variable of a given type, all within the body of a method with a given signature. They define a metric termed \emph{s-score}, which can be used to identify type usages that are themselves rare, but similar to a very common type usage. This parallels our goal of finding rare event-emitter pairs where a pair with a different event or a different emitter is very common, although the technical details of our approach are again somewhat more complex to deal with the problem of custom events.

A significant amount of research has been devoted to the
detection of \textit{event races} using static~\cite{zheng2011} and dynamic~\cite{petrov2012,raychev2013}
analysis. Recent work has focused on event races that have observable
effects~\cite{DBLP:conf/sigsoft/MutluTL15}, by classifying event races~\cite{zhang2017}, and by
developing specialized techniques focused on event races that occur during page
initialization~\cite{initracer2017} or that are associated with AJAX
requests~\cite{DBLP:conf/sigsoft/AdamsenMAT18}. The access paths used in this paper are not
precise enough to capture the ordering constraints necessary for event-race detection, so our
approach is not immediately applicable to this problem.

Other researchers have used statistical reasoning for predicting properties of programs for use in bug finding. 
Raychev et al. \cite{DBLP:conf/popl/RaychevVK15} derive probabilistic models from existing data
using structured prediction with conditional random fields (CRFs). They apply their analysis to
JavaScript programs to predict the names of identifiers and types of variables in new, unseen
programs, and suggest that the computed results can be useful for de-obfuscation and adding or
checking type annotations.
Eberhardt et al. \cite{DBLP:conf/pldi/EberhardtSRV19} apply unsupervised machine learning to a large
corpus of Java and Python programs obtained from public repositories to infer aliasing
specifications for popular APIs, which are then used to enhance a may-alias
analysis that is applied to applications using such APIs. 
The resulting enhanced analysis is demonstrated to lead to improvements in client
analysis such as typestate analysis (by eliminating a false positive result) and taint analysis (by
eliminating a false negative result).
Chibotaru et al. \cite{DBLP:conf/pldi/ChibotaruBRV19} present a semi-supervised method for inferring
taint analysis specifications. A propagation graph is inferred from each program
in a dataset, and it is assumed that a small number of nodes corresponding to API functions is
annotated as a source, sink, or sanitizer. To infer situations where unannotated nodes also play one of these roles,
a set of linear constraints
is derived from the propagation graph so that the solution to constraints represents the
likelihood of unannotated nodes being a source, sink, or sanitizer.
The program properties these works are designed to identify are API types and function signatures.
They do not discuss applications to message-passing systems like is seen in event-driven programming.

Hanam et al.~\cite{DBLP:conf/sigsoft/HanamBM16} present a technique for discovering
JavaScript bug patterns by analyzing many bug-fix commits. They decompose commits into a set
of language-construct changes, represent these as feature vectors, and apply unsupervised
machine learning to identify bug patterns. The identified
patterns are low-level issues such as dereferencing \code{undefined} and 
incorrect error handling. They do not discuss bug patterns related to event handling.

DeepBugs~\cite{DBLP:journals/pacmpl/PradelS18} aims to generate bug-fix
changes automatically. By applying simple program transformations to code
that is assumed to be correct, training data is obtained for a classifier that
distinguishes correct from anomalous code. The approach is evaluated for three types of
errors (swapped function arguments, wrong binary operator, wrong operand in binary operation),
and detected dozens of real bugs, with a false positive rate of
around 30\%. It is unclear how well this approach would work for less syntactic bugs like the
dead-listener bugs we consider.

Ryu et al.~\cite{ryu2018toward} present the SAFE tools for detecting type mismatch bugs that cause runtime errors (e.g., accesses to \code{undefined}) in JavaScript web applications.
They construct simple models of browser runtime constructs such as the HTML Document Object Model (DOM) through a dynamic analysis; this is used as input for their bug detector.
The SAFE tools differ from our work in three key ways: most importantly, the class of bugs SAFE tracks does not include dead-listener bugs; also, their target runtime is the browser while ours is Node.js; and, our analysis is purely static.

\section{Conclusion}\label{ConclusionSection}

We have presented an approach for detecting dead listener patterns in event-driven JavaScript programs that
relies on a combination of static analysis and statistical reasoning.
The static analysis computes a set of listener-registration pairs $\Pair{a}{e}$ where $a$ is an access path
and $e$ the name of an event, reflecting the fact that a listener is registered for $e$ on an object
represented by $a$.
After applying the static analysis to a large corpus of JavaScript applications, statistical modeling is used
to differentiate \LearnedCorrectTerm{} event listener registrations that are commonly observed from rarely observed 
\LearnedIncorrectTerm{} cases that 
are likely to be \ValidationSetIncorrectTerm{}.
In a large-scale evaluation on \TotalProjects{} open-source JavaScript code bases, our technique was able to detect \TotalIncorrectPairs{} \LearnedIncorrectTerm{} listener-registration patterns, while maintaining a precision of \BestTPRate{}\% and recall of \BestRecallRate{}\% over a validation set, demonstrating that a learning-based approach to
detecting event-handling bug patterns is feasible. 

We report on several additional experiments to better assess the impact of the data set analyzed by the 
statistical analysis, the utility of the results, and the practicality of the technique. 
One experiment revealed that the \textit{selection} of the particular subset of data that statistical 
analysis is trained on does not substantially affect the choice of optimal configuration. On the other hand, 
we found the \textit{size} of the subset used for training to have significant impact, with smaller
training set sizes generally resulting in classifiers that have unstable precision and lower recall on the
full data set.   
Furthermore, we demonstrated that our approach is effective at identifying buggy listener registrations
in real code bases: of the \NumReportedIssues{} issues we recently reported to developers of
\NumAnalyzedProjects{} open-source projects on GitHub, \NumConfirmedBugs{} were confirmed as bugs.
While the statistical analysis requires a significant amount of compute time, we would expect this cost
to be incurred infrequently, as APIs tend to evolve slowly. Checking a specific project for dead listeners
typically takes no more than a few minutes for all but the largest projects.

As future work, we plan to explore more precise notions of access paths that would allow us to build
distinct representations for function calls where some arguments are string literals
and others are callbacks.
In principle, this would enable us to distinguish access paths in the presence of nested event handlers.

\section*{Acknowledgment}
The authors would like to thank Albert Ziegler for insightful and helpful discussions about the statistical modeling. 

\noindent
E. Arteca and F. Tip were supported in part by National Science Foundation grants CCF-1715153 and CCF-1907727.
E. Arteca was also supported in part by the Natural Sciences and Engineering Research Council of Canada.


\bibliographystyle{IEEEtran}
\bibliography{IEEEabrv,paper}


\begin{IEEEbiography}[{\includegraphics[width=1in,height=1.25in,clip,keepaspectratio]{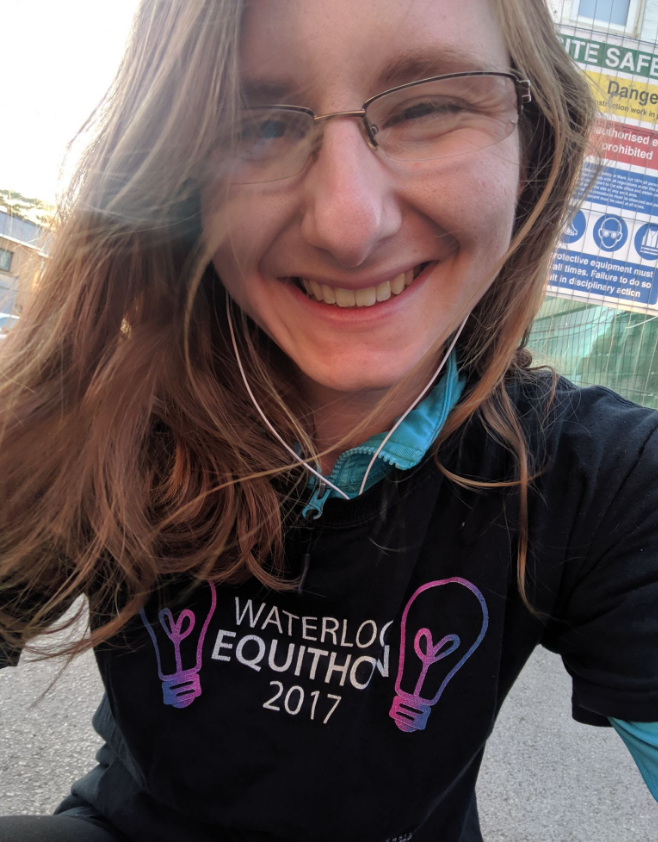}}]{Ellen Arteca} 
  is currently a PhD candidate at the Khoury College of Computer Sciences, Northeastern University, advised by Dr. Frank Tip. 
  She received her MMath degree from the University of Waterloo in 2018. 
  Her research interests include program analysis, bug finding, and test generation for dynamic languages.
\end{IEEEbiography}

\begin{IEEEbiography}[{\includegraphics[width=1in,height=1.25in,clip,keepaspectratio]{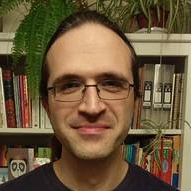}}]{Max Sch{\"{a}}fer} 
  received a DPhil in computer science from the University of Oxford in 2010. 
  He is currently a principal software engineer with GitHub in Oxford, UK. 
  His research interests include program analysis, software supply-chain security, and advanced programming tools.
\end{IEEEbiography}

\begin{IEEEbiography}[{\includegraphics[width=1in,height=1.25in,clip,keepaspectratio]{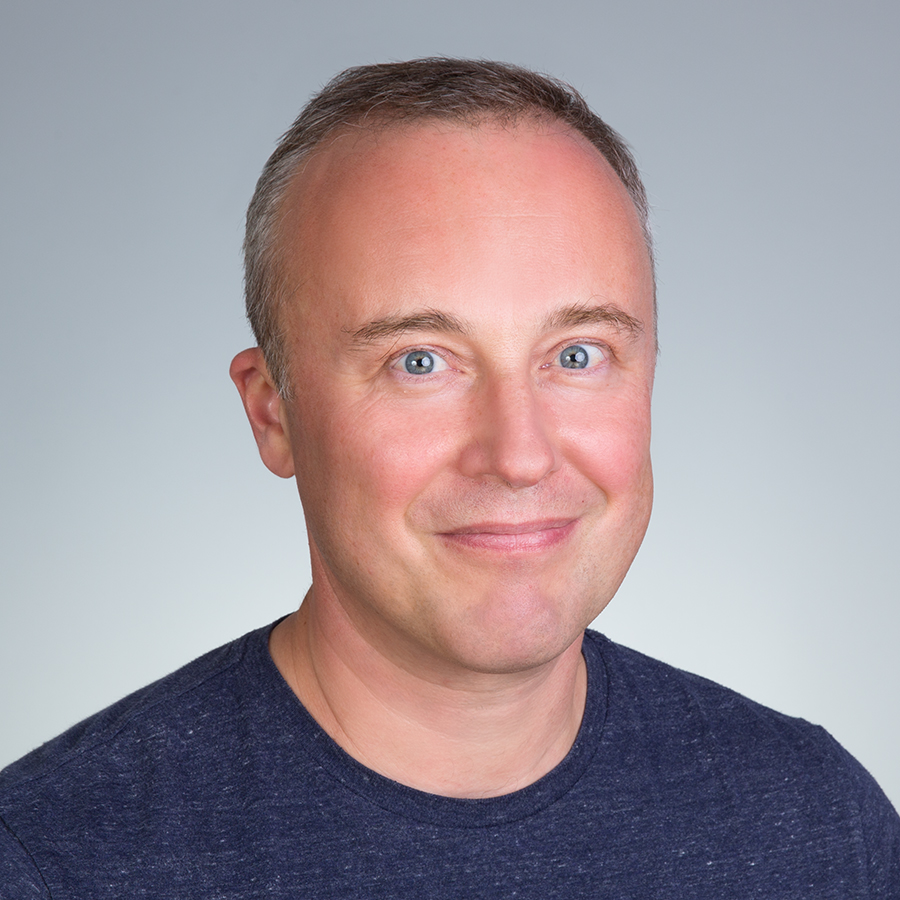}}]{Frank Tip} 
  received the PhD degree from the University of Amsterdam, Amsterdam, Netherlands, in 1995. 
  He is currently a professor at the Khoury College of Computer Sciences, Northeastern University, Boston, Massachusetts. 
  His research interests include 
  asynchronous Programming, static and dynamic program analysis, refactoring, and test generation.
\end{IEEEbiography}

\end{document}